\documentclass[review,3p,12pt]{elsarticle}

\usepackage{color}
\usepackage{mathtools}
\usepackage{hyperref} 
\usepackage{setspace}

\biboptions{numbers,compress}

\journal{Applied Mathematics and Computations}

\begin{document}

\begin{frontmatter}

\title{\textsc{Characterization of micro-Capsules Deformation in Branching Channels}}

\author[dei]{A. Coclite\corref{cor}}
\ead{alessandro.coclite@poliba.it}

\author[dmmm]{M. D. de Tullio}
\ead{marcodonato.detullio@poliba.it}

\author[dmmm]{G. Pascazio}
\ead{giuseppe.pascazio@poliba.it}

\author[dei]{T. Politi}
\ead{tiziano.politi@poliba.it}

\cortext[cor]{Corresponding author}

\address[dei]{Dipartimento di Ingegneria Elettrica e dell'Informazione, Politecnico di Bari, Via Re David 200 -- 70125 Bari, Italy}

\address[dmmm]{Dipartimento di Meccanica, Matematica e Management, Politecnico di Bari, Via Re David 200 -- 70125 Bari, Italy}

\begin{abstract}

In this paper, the dynamic of inertial capsules into microfluidic bifurcations is studied. The fluid evolution is based on the solution of the BGK -- lattice Boltzmann scheme including a forcing term accounting for immersed geometries. The dynamic-Immersed Boundary forcing strategy is adopted for imposing no-slip boundary conditions on moving deformable or rigid structures, while, on fixed immersed geometries the Bouzidi-Firdaouss-Lallemand second-order bounce back technique is implemented. The proposed computational framework is employed to detail dynamics and deformation of rigid and deformable capsules traveling into a branching duct. This journey is characterized in terms of i) the capsule/bifurcation interaction depending on the sharpness of the branching channels junction; ii) daughter branches aperture angle; iii) occlusion ratio, the ratio between capsule size and main channel diameter; iv) flowing capsules stiffness; v) number of flowing particles. 

\end{abstract}
\end{frontmatter}

\section*{Introduction}

As widely known, the micro-circulatory network can be assimilated to a collection of branching channels in a self-similarity paradigm up to the micrometric scale~\cite{guidolin2011}. In this context, the transport of a particle dragged by an external fluid through a branching microchannel is a prototypical problem that still needs to be thoroughly addressed. The importance of this fundamental problem resides in the large number of physical phenomena that can be enlightened depending on the established flow and navigating particles' characteristics. Such phenomena need to be accurately accounted for describing the nonlinear behavior of microfluidic devices, such as those used in particle sorting lab-on-chip\cite{dicarlo2019,volpe2017}. 
For example, for moderate Reynolds numbers, the path selected during the journey can be predicted for simple channel geometries as a function of the particle releasing position and inertia~\cite{koolivand2017,gandhi2020,tran2020}. On the other side, the transport of soft particles in branching ducts is also considered for the rational design of biomedical devices able to enrich the concentration of micro-objects with specific characteristics as a result of the sorting induced by the established flow~\cite{doyeux2011,sevenie2015,wang2018,coclite20194,lu2021}. Specifically, one can think to focus on the altered mechanical properties and dynamics due to the altered structure of the cytoskeleton in diseased cells for the design of such sorting devices~\cite{hymel2019,haner2020}. Indeed, such separation techniques, usually referred as label-free, do not restrict to cells and are largely used for capsules and particles sorting in microfluidics chip~\cite{shields2015,vesperini2017,haner2021,guzniczak2020}. Recently, the microdevices design community is pushing toward the idea of mimicking the authentic complexity of nature with \textit{in-vitro} models for extravascular mass transport and vascular transport~\cite{manneschi2016,miali2019,mollica2019}. As a consequence, the scientific community invested more and more efforts in reliable computational tools to model the fluid-structure interaction of micrometric constructs navigating complex geometries mimicking the circulatory system~\cite{melchionna2011,freundARFM2014,ye2014,secombARFM2017,sebastian-dittrichARFM2018,coclite20202,coclite20203}. All of the considered applications rely on the specific modeling of the physical phenomena involved and, thus, the computational tools aiming for the prediction of such phenomena must be formulated establishing a compromise between accuracy and affordability of necessary computations.

The present paper aims to discuss the transport and to measure the deformation of inertial capsules with different stiffness and sizes navigating symmetric branching vessels as a function of the bifurcation geometry and particles characteristics. In particular, the authors thoroughly detail the strain experienced by such capsules underlying the parameters ranges for which large as well as minimal or negligible deformations are measured. To this scope, a combined Lattice-Boltzmann / dynamic-Immersed--Boundary (IB) method~\cite{coclite20191,coclite20204} is employed due to its ability to predict the transport of massive capsules within arbitrarily shaped immersed geometries. Specifically, immersed objects are described by their surface considered as an ensemble of Lagrangian vertices composing a triangular mesh that moves on the Eulerian grid used to solve the flow equations. The fluid evolution is obtained through a BGK-lattice Boltzmann scheme for the Navier-Stokes equation in the incompressible regime. Dirichlet Boundary Conditions (BC) are enforced on external boundaries adopting the Zou-He bounce back procedure~\cite{zouhe1997}; while, on immersed objects, the BCs are imposed in two different fashions depending on whether they move or not. For fixed obstacles, the Bouzidi-Firdaouss-Lallemand technique is adopted to bounce back the distribution functions belonging to surrounding fluid nodes and incoming on solid nodes~\cite{bouzidi2001}. On moving geometries, BCs are enforced by projecting the forcing term computed into the dynamic-IB framework onto the computational molecula. Structures transport is readily obtained by solving the equation of motion along with the constitutive model adopted for particle deformations~\cite{coclite20204}. The proposed model is implemented to detail the transport of inertial capsules slightly denser ($\simeq 10\%$) than the surrounding fluid within a circular cross-section branching channel. This analysis is carried out systematically by studying the journey of such capsules in terms of i) capsule/bifurcation interaction and sharpness of the bifurcation junction and - two different junctions are proposed and analyzed; ii) symmetric daughter branches aperture angle concerning the main channel axis $\theta$ (= $\pi/6$, $\pi/3$, and $\pi/2$); iii) occlusion ratio, the ratio between capsule size $d_p$ and main channel diameter $d$ ($d_p/d$ = 1/3, 1/2, 2/3, and 5/6); iv) flowing capsules mechanical stiffness, regulated by  the parameter Ca corresponding to the relative effect of hydrodynamics drag over capsule's strain resistance - five different values for Ca are considered (Ca = 0.01, 0.025, 0.05, 0.075, and 0.1); v) number of flowing particles - single particle as well as multiple particle transport is considered for soft and rigid capsules. 

\section{Numerical method and computational procedure}

In this section a brief description of the dynamic-Immersed–Boundary method combined with a BGK-Lattice–Boltzmann technique, developed and extensively validated by the authors is given~\cite{coclite20191,coclite20204}. The fluid evolution is obtained on a three-dimensional lattice while the immersed body surface is modeled as a collection of Lagrangian points subject to a constitutive equation.

\subsection{\textsc{Fluid evolution}}

\begin{figure}[h!]
\centering
\includegraphics[scale=0.2]{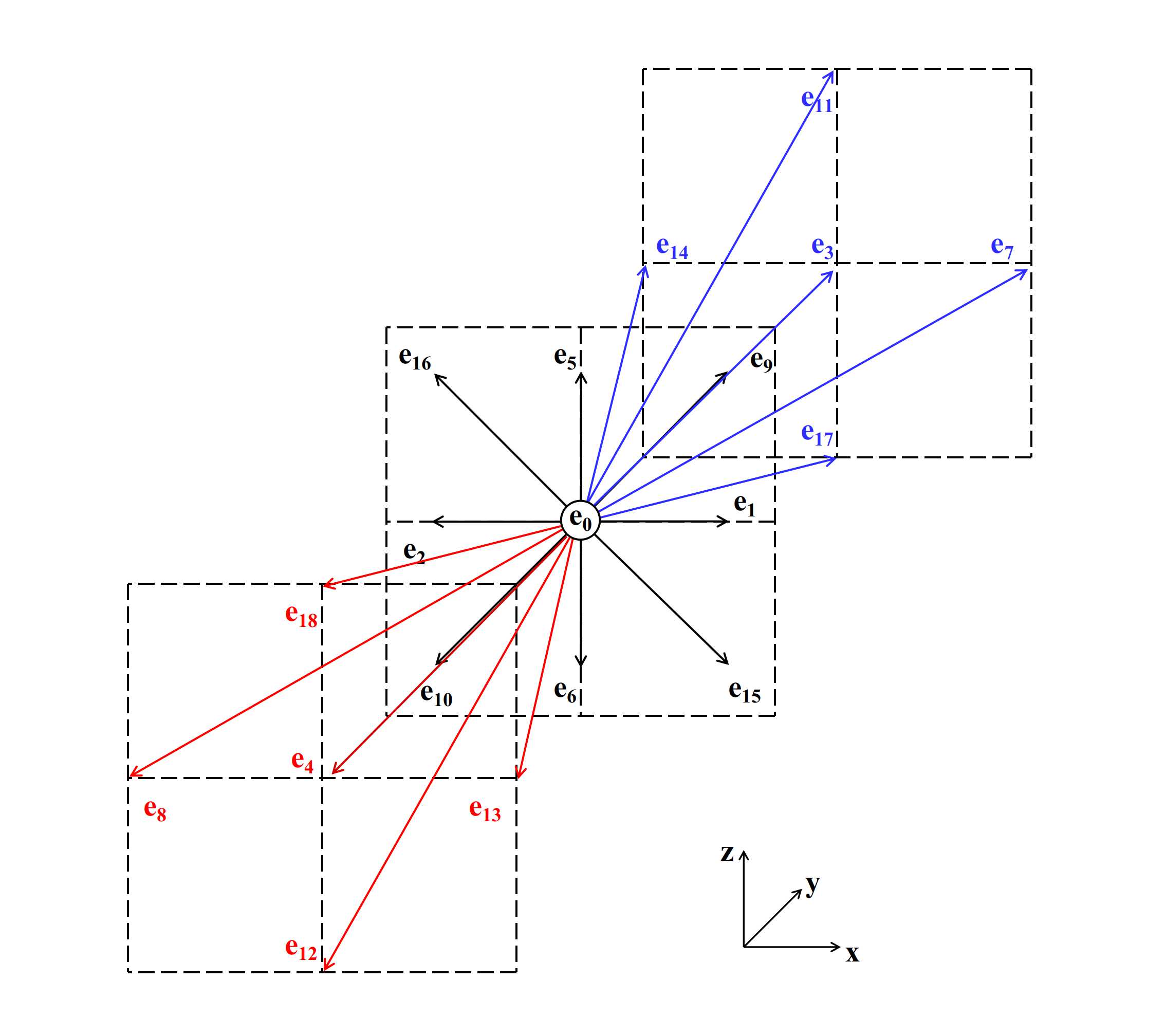}
\caption{{\bf The D3Q19 lattice molecule.} Definition of the D3Q19 computational molecule with ${\bf e}_i$ vectors as defined in {\bf Eq.\eqref{e_i}} (not in scale).}
\label{D3Q19}
\end{figure}
The incompressible Navier-Stokes equation for the fluid evolution is recovered with second order accuracy through the BGK-approximation of the Boltzmann equation on a three-dimensional manifold~\citep{BGK}. Such equation is solved onto the velocities space on a 3D molecule with 19 reticular velocities namely, $\{ {\bf e}_i({\bf x}) \}_{i=0,\dots,18}$ where ${\bf x}=(x,y,z)\hat{\bf x}$ is the coordinates vector (see {\bf Figure.\ref{D3Q19}}):\\
\begin{small}
\begin{equation}
{\bf e} = \begin{pmatrix}
0& 1&-1& 0& 0& 0& 0& 1& 1& 1& 1&-1&-1&-1&-1& 0& 0& 0& 0 \\
0& 0& 0& 1&-1& 0& 0& 1&-1& 0& 0& 1&-1& 0& 0& 1& 1&-1&-1 \\
0& 0& 0& 0& 0& 1&-1& 0& 0& 1&-1& 0& 0& 1&-1& 1&-1& 1&-1 \\
\end{pmatrix}\, .
\label{e_i}
\end{equation}
\end{small}
The solution of the Boltzmann equation is obtained in terms of a vector in the velocity space composed by 19 discrete functions ${\bf f} = \{f_i ({\bf x})\}_{i=0,\dots,18}$ evolving on this computational molecula,
\begin{equation}
\label{evoleq}
{f_{i}({\bf x}+{\bf e}_{i}\Delta t, t+\Delta t)-f_{i}({\bf x}, t)=
-\frac{\Delta t}{\tau}[f_{i}({\bf x}, t)-f_{i}^{eq}({\bf x}, t)] +\Delta t {\cal F}_i}\, ,
\end{equation}
where $\tau$ is the relaxation time, $\Delta t$ the time step, $\{f^{eq}_i ({\bf x})\}_{i=0,\dots,18}$ the components of the equilibrium discrete functions while ${\cal F}_i$ the ones of the discrete forcing term. The kinematic viscosity $\nu$ is obtained as a function of $\tau$, $\nu=c_s^2(\tau -\frac{\Delta t}{2})$ being $c_s=1/\sqrt{3}$.
Macroscopic thermodynamics quantities are obtained as statistical moments of ${\bf f}$: 
\begin{itemize}
\item the fluid density $\rho=\sum_i f_i$;
\item the momentum $\rho {\bf u}=\sum_i f_i {\bf e}_i+\frac{\Delta t}{2}{\bf f}_{ib}$;
\item the thermodynamics pressure $p= c_s^2 \sum_i f_i = c_s^2 \rho$.  
\end{itemize}
To recover the Navier-Stokes equation, the vector ${\bf f^{eq}}$ is defined as the expansion of the Maxwell-Boltzmann distribution,
\begin{equation}
\label{eqfunc}
f_i^{eq}({\bf x},t)=\omega_i\rho\left[1+\frac{1}{c_s^2}({\bf e}_i \cdot{\bf u})+
\frac{1}{2 c_s^4}({\bf e}_i \cdot{\bf u})^2-\frac{1}{2c_s^2}{\bf u}^2
\right]\, ,
\end{equation}
with $\omega_0=1/3$, $\omega_{1-6}=1/18$, and $\omega_{7-18}=1/36$.
In the same fashion, the expansion of ${\cal F}_i$ reads~\cite{guo2002,coclite2014}:
\begin{equation}
\label{forcing}
{{\cal F}_i= \Bigl(1-\frac{1}{2\, \tau}\Bigr)\omega_i\Bigl[\frac{{\bf e_i}-{\bf u}}{c_s^2}+\frac{{\bf e_i}\cdot{\bf u}}{c_s^4}{\bf e_i}\Bigr]\cdot {\bf f}_{ib}}\, 
\end{equation}
where ${\bf f}_{ib}({\bf x},t)$ is the force term due to the presence of immersed structures~\cite{coclite20163,coclite20191}.

\subsection{\textsc{Dirichlet conditions for external boundaries}}

Dirichlet Boundary Conditions (DBC) are enforced by adopting the bounce back procedure by Zou and He~\cite{zouhe1997} to enforce velocity- or pressure-based boundary conditions. As widely known {\bf Eq.\eqref{evoleq}} is solved numerically in two steps, the collision, and the streaming phases. However, for boundaries, the streaming phase cannot be completed since it is not possible to stream in and out the distribution functions for the boundary nodes. Let us consider, for example, the north wall ($z = z_{max}$). All of the distribution functions entering the wall are known since they derive from the post-collision streaming phase of the nodes belonging from the plane $z = z_{maz}-\Delta x$, namely $f_5, f_9,\, f_{11},\, f_{16}$ and $f_{18}$ (see {\bf Figure.\ref{D3Q19}}). On the contrary, all of the distribution functions that should be streamed back into the wall nodes from the plane $z = z_{max}+\Delta x$, namely $f_6, f_{10},\, f_{12},\, f_{15}$ and $f_{17}$ (see {\bf Figure.\ref{D3Q19}}) are unknown and need to be somehow imposed or computed. Zou and He proposed to enforce conservative boundary conditions through a sophisticated bounce-back procedure in which the unknown distribution functions are computed also by imposing the value of three macroscopic quantities, namely, the three components of the velocity field for velocity-based DBCs or the pressure and the two non-orthogonal components (with respect to the boundary normal) of the velocity for pressure-based DBCs. 
Therefore, for the north wall the velocity-based DBCs read:
{\small \begin{equation}
{\begin{cases}
\rho(x_w,y_w,z_w) = (u_{z,w}+1)^{-1}\Bigl[f_0+f_1+f_2+f_3+f_4+f_7+f_8+f_{13}+f_{14}+ 2 \Bigl( f_5+f_9+f_{11}+f_{16}+f_{18}\Bigr) \Bigr]\\
f_6 = f_5-\frac{1}{3}\rho(x_w,y_w,z_w)u_{z,w}\\
f_{15} = f_{16}+\frac{1}{6}(-u_{z,w}+u_{x,w})-N_{z,x}\\
f_{10} = f_9+\frac{1}{6}(-u_{z,w}-u_{x,w})+N_{z,x}\\
f_{17} = f_{18}+\frac{1}{6}(-u_{z,w}+u_{y,w})-N_{z,y}\\
f_{12} = f_{11}+\frac{1}{6}(-u_{z,w}-u_{y,w})+N_{z,y}\\
\end{cases}}\, ,
\label{extbound}
\end{equation}}
where $u_{x,w}$, $u_{y,w}$, and $u_{z,w}$ are the values of the velocity components imposed at the wall nodes and 
\begin{eqnarray}
N_{z,x} &=& \frac{1}{2}\Bigl( f_1+f_7+f_{13}-(f_2+f_{14}+f_8)\Bigr) - \frac{1}{3}\rho(x_w,y_w,z_w)u_{x,w}\, , \\
N_{z,y} &=& \frac{1}{2}\Bigl( f_3+f_7+f_{14}-(f_4+f_{13}+f_8)\Bigr) - \frac{1}{3}\rho(x_w,y_w,z_w)u_{y,w}\, .
\end{eqnarray} 

\subsection{\textsc{No-slip conditions for moving immersed boundaries}}

\begin{figure}[h!]
\centering
\includegraphics[scale=0.2]{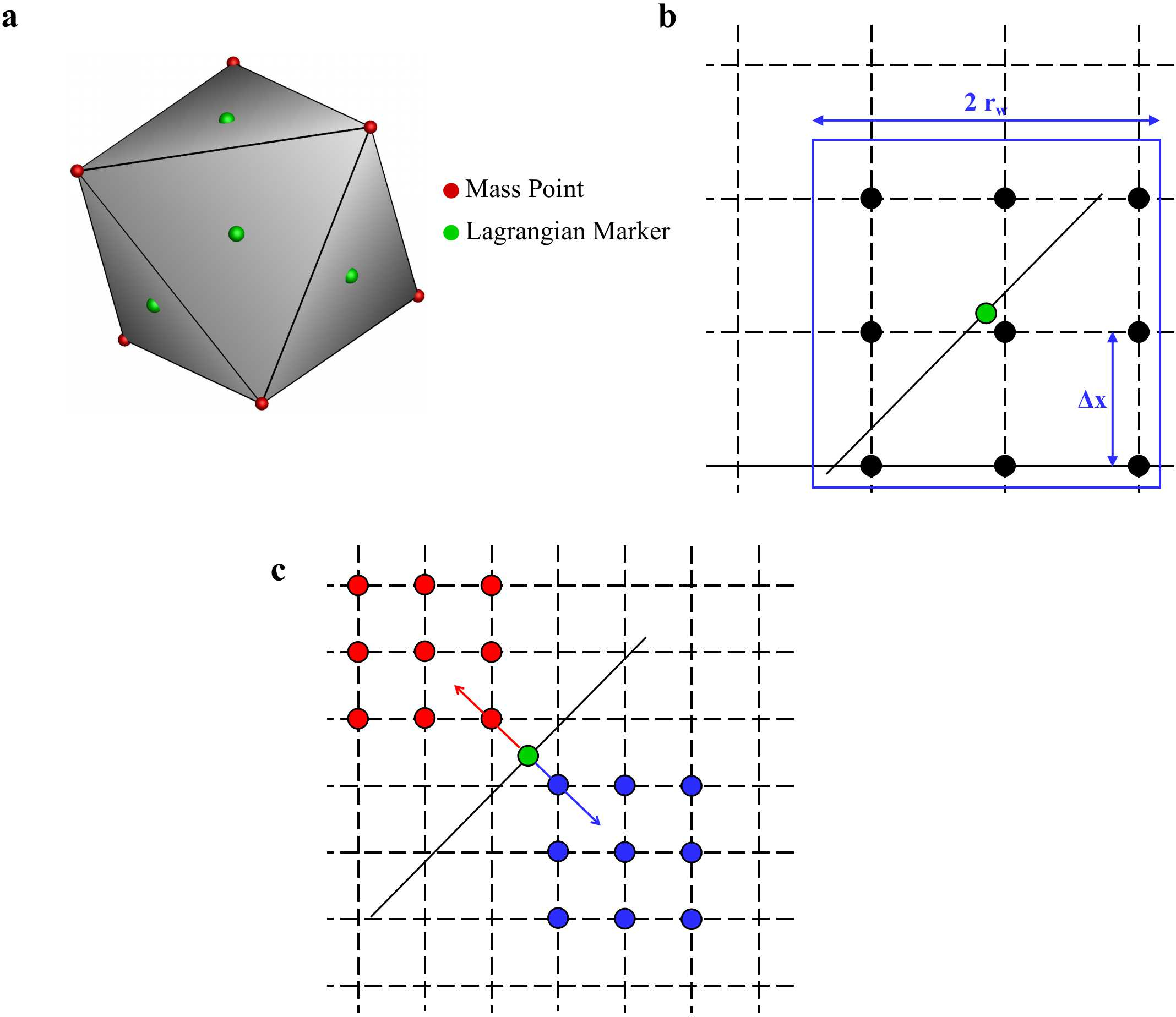}
\caption{{\bf Immersed-boundary technique schematic.} {\bf a.} Spatial approximation of an immersed body obtained with a triangular mesh. Each element centroid corresponds to a Lagrangian marker. {\bf b.} Eulerian points (per plane) on which the dynamic-IB forcing procedure is obtained through the support (blue square) centered on the selected Lagrangian marker (green circle). {\bf c.} Eulerian points (per plane) used for hydrodynamics stresses computations taken at 1.2 $\Delta x$ along the inward and outward normal directions of each element.}
\label{IB}
\end{figure}

In the Immersed-Boundary technique~\cite{peskin2002}, the surface of an arbitrarily shaped body immersed in the flow is modeled as an ensemble of Lagrangian vertices moving on the withstanding Eulerian grid on which the fluid solution is computed ({\bf Figure.\ref{IB}a}). Here, such surface is discretized utilizing a triangular mesh and the centroids of each element are called ``Lagrangian markers''. A convenient transfer function is built on a moving least squares reconstruction for exchanging needed information between the two meshes~\cite{coclite20191,coclite20204,MDdTJCP2016}.
Specifically, for each triangular element centroid, 9 Eulerian points per plane are identified as all of the Eulerian points falling into the support domain by side equal to $2 r_w = 2.6\, \Delta x$ and centered on the selected centroid (indexed here with $l$); $\Delta x$ being the Eulerian grid spatial resolution (see {\bf Figure.\ref{IB}b}). Given the fluid solution at time step $n$, the velocity the marker $l$ is approximated by,
\begin{equation}
\label{uLag}
{U_i({\bf x})=\sum_{k=1}^{27}\, \phi^l_k({\bf x}) u^k_i}\, ,
\end{equation}
being $u^k_i$ the component $i$ of the fluid velocity field ${\bf u}$ taken at the Eulerian point $k$ associated to the marker $l$. $\phi$ is the operator given by the minimization with respect to ${\bf a(x)}$ of:
\begin{equation}
\label{l2norm}
{J=\sum_{k=1}^{27} W({\bf x}-{\bf x}^k}) [{\bf p}^T({\bf x}^k){\bf a(x)}-u^k_i]^2\, .
\end{equation}
In the latter, ${\bf x}$ is the coordinate vector of the marker, ${\bf p}^T$ is a linear basis vector, ${\bf a(x)}$ a vector of coefficients such that 
$\sum_{k=1}^{27} \, \phi^l_k({\bf x}) u^k_i\, ={\bf p}^T({\bf x}^k){\bf a(x)}$. 
$W({\bf x}-{\bf x}^k)$ is the weight function:
\begin{equation}
\label{intKernel}
W({\bf x}-{\bf x}^k)= 
\begin{cases} 
e^{-(r_k/\alpha)^2}\,  &\mbox{if } r_k\leq 1 \\ 
0\,  & \mbox{if } r_k> 1\ \\
\end{cases} \, , 
\end{equation} 
where, $\alpha=0.3$ and $r_k$ is the distance between the marker and point $k$ normalized with respect to the support domain size length.
For each Lagrangian marker, the volume force ${\bf F}_l$ required to enforce no-slip BCs is computed, and then transferred to the Eulerian points falling into the conveniently defined support domain.
Once the velocity at the $l$-th Lagrangian marker is known from Eq.~\ref{uLag}, the desired velocity at that point, ${\bf U}_{b}^l({\bf x})$, is enforced by the volume force:
\begin{equation}
{{\bf F}_l {\bf (x)}=\frac{{\bf U}_{b}^{l} {\bf (x)}-{\bf U(x)}}{\Delta t}}\, ,
\end{equation}
that is then used to compute the force for the point $k$,
\begin{equation}
\label{bodyforce}
{{\bf f}_{ib}}^k=\sum_{l} c_l \phi^l_k {\bf F}_l \, .
\end{equation}
Note that, in the above equation the sum runs over all markers having the point $k$ point in their support domain. $c_l$ is a coefficient obtained by imposing the conservation of the total force exerted by the fluid~\cite{vanella2009} between the two grids.

\subsection{\textsc{Immersed-Structure Transport}}

Immersed bodies motion due to the interaction established with the fluid is modeled by adopting the \textit{dynamics-IB} technique described in \cite{coclite20191,coclite20204}. For the case of soft objects, the equation of motion of each Lagrangian point is solved considering both constitutive model response and hydrodynamics forces. The case of rigid bodies is treated separately by integrating Newtown's second law and the moment equation for each body.

\textbf{\textit{Deforming bodies constitutive model.}} Deforming bodies internal forces are computed as elastic strain response, bending resistance, and enclosed volume conservation~\cite{coclite20204}. The stretching elastic potential acting on the two vertices sharing the $l$-th edge is given by
\begin{equation}
{V_{l}^{s}=\frac{1}{2}k_{s,l}(l_{l}-l_{l,0})^{2}}\, ,
\label{strainPot}
\end{equation}
being $k_{s,l}$ [N/m] the elastic modulus for the edge $l$, $l_{l}$ its length in the deformed configuration and $l_{l,0}$ its length at the resting. The elastic forces for vertices 1 and 2 connected by the edge \textit{l} are:
\begin{equation}
{\begin{cases}
{\bf F}_{1}^{int,s}=-k_{s,l}(l-l_{0})\frac{{\bf r}_{1,2}}{l}\,  \\
{\bf F}_{2}^{int,s}=-k_{s,l}(l-l_{0})\frac{{\bf r}_{2,1}}{l}\,  \\
\end{cases}}\, ,
\label{strainFor}
\end{equation}
where ${\bf r}_{i,j}={\bf r}_{i}-{\bf r}_{j}$ and ${\bf r}_{i}$ is the ray vector connecting \textit{i} to \textit{j}. The elastic modulus of the immersed surface is obtained trough the nondimensional group measuring the ratio between viscous drag forces versus surface tension forces, the capillary number Ca$=\frac{\rho \nu u_{ref}}{k_{s,0}}$. Then, it is spread to all edges considering the fraction of the total surface assigned to each edge, namely the sum of the area of the two elements sharing the selected edge $A^l_i$ : $k_{s,l}=\frac{k_{s,0} T \sum_i A^l_i}{l^2}$~\cite{gelder}. 
Indeed, $\rho$ and $\nu$ are the fluid density and kinematic viscosity; $u_{ref}$ the reference velocity, $k_{s,0}$ the surface elastic modulus, and $T$ the surface thickness.

Off-plane deformations are constrained by a bending potential depending on the difference between the local curvature in the resting and the deformed configurations, $c_{v,0}$ and $c_{v}$, respectively,
\begin{equation}
{V_{v}^{b}=\frac{1}{2}k_{b}(c_{v}-c_{v,0})^{2}}\, ,
\label{bendPot}
\end{equation}
being $k_{b}$ the bending modulus. $c_{v}$ is computed through the angle defined by the unit normal vectors of two adjacent elements, ${\bf n}_1$ and ${\bf n}_2$, ($\theta -\theta _{0}$), with $\theta _{0}$ the angle in the resting configuration. 
$k_{b}$ is obtained as a function of $k_{s,l}$ through the nondimensional group $Eb=\frac{k_{b}}{k_{s,l}r^2}$.~~\cite{tan2012}  
The resulting force on the involved four nodes reads:
\begin{equation}
{\begin{cases}
{\bf F}^{int,b}_1 = \beta_b \bigl[
b_{11}({\bf n}_1 \times {\bf r}_{32}) + 
b_{12}({\bf n}_2 \times {\bf r}_{32}) \bigr] \, \\
{\bf F}^{int,b}_2 = \beta_b \bigl[
b_{11}({\bf n}_1 \times {\bf r}_{13}) + 
b_{12}({\bf n}_1 \times {\bf r}_{34}  + {\bf n}_2 \times {\bf r}_{13}) +
b_{22}({\bf n}_2 \times {\bf r}_{34}) \bigr] \, \\
{\bf F}^{int,b}_3 = \beta_b \bigl[
b_{11}({\bf n}_1 \times {\bf r}_{21}) + 
b_{12}({\bf n}_1 \times {\bf r}_{42}  + {\bf n}_2 \times {\bf r}_{21}) +
b_{22}({\bf n}_2 \times {\bf r}_{42}) \bigr] \, \\
{\bf F}^{int,b}_4 = \beta_b \bigl[
b_{11}({\bf n}_1 \times {\bf r}_{23}) + 
b_{22}({\bf n}_2 \times {\bf r}_{23}) \bigr] \, \\
\end{cases}}\, ,
\label{bendFor}
\end{equation}
with $b_{ij}= \frac{{\bf n}_i\cdot {\bf n}_j}{|{\bf n}_i||{\bf n}_j|}$ and $\beta_b=k_b\frac{sin(\theta)cos(\theta_0)-cos(\theta)sin(\theta_0)}{\sqrt{1-cos^2(\theta)}}$.

The incompressibility (conservation of the total volume) of the membrane is enforced through a penalty model:
\begin{equation}
{{\bf F}_{i}^{int,v}=-k_{v}\left(1-\frac{V}{V_{0}}\right) {\bf n}_{i} A_{i}}\, , 
\label{volFor}
\end{equation}
$k_v$ an incompressibility constant; $V$ and $V_0$ the current and initial volume enclosed by the structure; $A_i$ the area associated with the element $i$.

\textbf{\textit{Hydrodynamics stresses.}} The hydrodynamic force acting on an element is composed by two terms, ${\bf F}_{i}^{ext,p}(t)$ and ${\bf F}_{i}^{ext,\tau}(t)$:
\begin{equation}
{{\bf F}_{i}^{ext,p}(t)=-(p^+_{i}-p^-_{i}) \, {\bf n}^+_{i} \, A_{i}}\, , \\
\label{hydroForP}
\end{equation}
\begin{equation}
{{\bf F}_{i}^{ext,\tau}(t)=(\bar{\bf \tau}^+_{i}-\bar{\bf \tau}^-_{i})\cdot {\bf n}^+_{i} \, A_{i}}\, .
\label{hydroForNu}
\end{equation}
$p^+_{i}$, $p^-_{i}$ and $\bar{\bf \tau}^+_{i}$, $\bar{\bf \tau}^-_{i}$ are the pressure and the viscous stress tensor taken on the marker associated to element $i$. Such, quantities accounts for both external (+) and internal (-) flow fields. ${\bf n^+}_{i}$ is the normal outward unit vector. Pressure and velocity derivatives in {\bf Eq.s \eqref{hydroForP}} and {\bf .\eqref{hydroForNu}} are computed at a distance 1.2 $\Delta x$ along the outward (+) and inward (-) normal directions (see {\bf Figure.\ref{IB}c})~\cite{MDdTJCP2016}.

\textbf{\textit{Bodies interaction and collision treatment.}} The interaction between two or more immersed bodies is treated by imposing a purely repulsive potential centered in each Lagrangian marker of each immersed object. Such repulsive force is such that the minimum distance allowed between two Lagrangian markers belonging to two different immersed structures is $\Delta x$. Specifically, the force exerted by the $i$-th Lagrangian marker of a particle at a distance $d_{i,j}$ from the $j$-th marker of a second particle is chosen to be proportional to the module of the characteristic velocity of the flow $u_{max}$:
\begin{equation}
{{\bf F}_i^{ext,c} = \frac{u_{max} \Delta t}{8\sqrt{2}}\sqrt{\frac{\Delta x}{d_{i,j}^5}}{\bf n}^-_i}\, .
\label{collision}
\end{equation} 

\textbf{\textit{Soft structure motion.}} The position of each element centroid is updated in time by integrating the equation of motion:
\begin{equation}
{m_{v}\dot{{\bf u}}_{v}={\bf F}_{v}^{int,s}(t)+{\bf F}_{v}^{int,b}(t)+{\bf F}_{v}^{int,v}(t)+{\bf F}_{v}^{ext,c}(t)+{\bf F}_{v}^{ext,p}(t)+{\bf F}_{v}^{ext,\tau}(t)}\, ,
\label{newtonSoft}
\end{equation}
where $m_v$ is the mass associated to the point $v$ with the assumption that the structure mass is uniformly distributed over the vertices; $\dot{{\bf u}}_{v}$ corresponds to the Lagrangian point acceleration. Such equation is integrated in time adopting the Verlet algorithm with the zero-step tentative velocity ${\bf u}_{v,0}(t+\Delta t)$ obtained by interpolating the fluid velocity from the Eulerian points falling in the support domain centered in $v$, 
\begin{equation}
{{\bf x}_{v}(t+\Delta t)=
{\bf x}_{v}(t)
+{\bf u}_{v,0}(t+\Delta t)
\Delta t+\frac{1}{2}\frac{{\bf F}_{v}^{tot}(t)}{m_{v}}\Delta t^{2}+O(\Delta t^{3})}\, ;
\label{verletPos}
\end{equation}
then the vertex velocity is computed as:
\begin{equation}
{{\bf u}_{v}(t+\Delta t)=\frac{\frac{3}{2}{\bf x}_{v}(t+\Delta t)-2{\bf x}_{v}(t)+\frac{1}{2}{\bf x}_{v}(t-\Delta t)}{\Delta t}+O(\Delta t^{2})}\, ,
\label{verletVel}
\end{equation}
and ${\bf x}_{v}(t+\Delta t)$ is finally corrected:.
\begin{equation}
{{\bf x}_{v}(t+\Delta t)=
{\bf x}_{v}(t)
+{\bf u}_{v}(t+\Delta t)
\Delta t+\frac{1}{2}\frac{{\bf F}_{v}^{tot}(t)}{m_{v}}\Delta t^{2}+O(\Delta t^{3})}\, .
\label{verletPosC}
\end{equation}

\textbf{\textit{Rigid structures transport.}} The transport of rigid structures is obtained by summing all force contributions over the structure surface ({\bf Eq.s.\eqref{hydroForP}}, {\bf .\eqref{hydroForNu}} and {\bf .\eqref{collision}}) and then computing the linear and angular velocities and positions. The Newton-Euler system reads,
\begin{equation}
\begin{cases}
{\bf F}^{tot}(t)=m\, \dot{{\bf u}}(t)\\
{\bf M}^{tot}(t)={\bf I}\, \dot{{\bf \omega}}(t)
\end{cases}\, .
\end{equation}
${\bf F}^{tot}(t)$ and ${\bf M}^{tot}(t)$ are the total force and moment; $m$ is the total structure mass; ${\bf I}$ the inertia tensor. ${\bf u}(t)$ and ${\bf \omega} (t)$ are derived as:
\begin{equation}
\begin{cases}
{\bf u}(t)=\frac{2}{3}(2{\bf u}(t-\Delta t)-\frac{1}{2}{\bf u}(t-2\Delta t)+\dot{{\bf u}}(t)\Delta t)+O(\Delta t^{2})\\
{\bf \omega} (t)=\frac{2}{3}(2{\bf \omega} (t-\Delta t)-\frac{1}{2}{\bf \omega} 
(t-2\Delta t)+\dot{{\bf \omega}}(t)\Delta t)+O(\Delta t^{2})\end{cases}\, .
\label{angularVelRig}
\end{equation} 

\subsection{\textsc{No-Slip boundary conditions for stationary immersed boundaries}}

\begin{figure}[h!]
\centering
\includegraphics[scale=0.2]{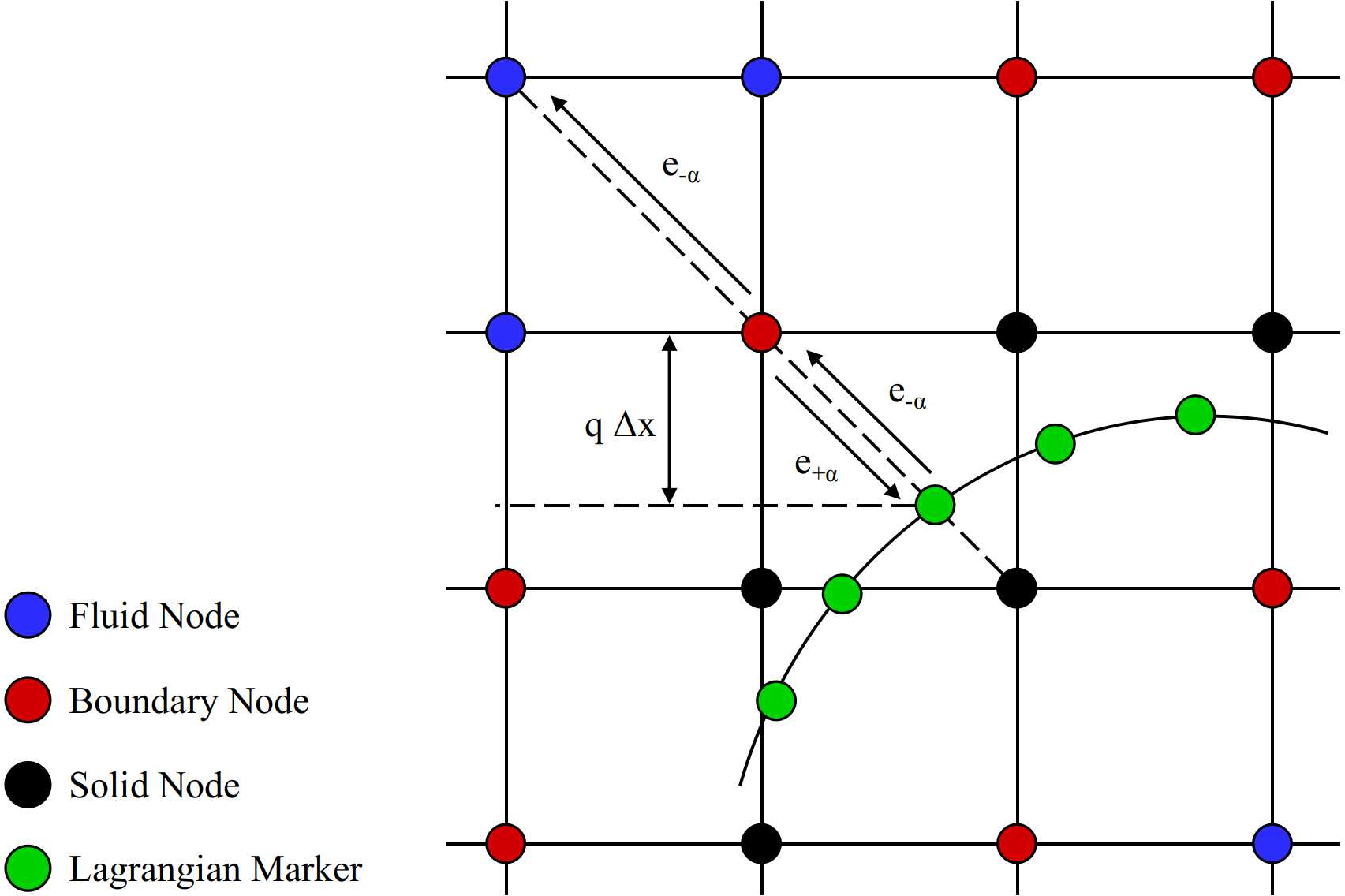}
\caption{{\bf Bouzidi-Firdaous-Lallemand interpolation technique.} Definition of solid and boundary nodes involved into the enforcement of the boundary conditions along with the distance q used for the interpolation of the bounced-back distribution functions.}
\label{Bouzidi}
\end{figure}
To limit the computational burden and save computational time, boundary conditions for stationary bodies and moving geometries are treated with two different techniques. Specifically, the second-order Bouzidi-Firdaouss-Lallemand technique is adopted to bounce-back distribution functions belonging from surrounding computational nodes incoming against the computational nodes occupied by the solid obstacle~\cite{bouzidi2001}. In the same fashion used for moving geometries, in this context, an immersed body corresponds to a collection of triangular elements. All of the computational nodes are tagged as ``fluid'', ``solid'' or ``boundary''. Eulerian points corresponding to computational cells in which a Lagrangian marker falls are considered ``solid nodes''. Then, a node into the computational domain not tagged as solid that is at a distance ${\bf e}_i\Delta t,\, i=0, ..., 18$ from a solid node is tagged as ``boundary'' node, being ${\bf e}_i$ the 19 reticular directions, all other nodes are considered as ``fluid nodes'' (see {\bf Figure.\ref{Bouzidi}}). Note that, within this parametrization, boundary nodes play a major role being the nodes on which the no-slip boundary conditions are enforced. Once the computational domain is completely tagged, for all of the boundary nodes are identified the reticular directions along which at least a solid point is found and unity value is assigned to the scalar field $a_i({\bf x})$ otherwise null (see {\bf Figure.\ref{Bouzidi}}). Moreover, for such directions, the normalized distance $q_i({\bf x})$ is computed as the ratio between the distance from the corresponding Lagrangian marker (with coordinates $({\bf x}_L)$) and the boundary node (with coordinates $({\bf x}_b)$) over the distance between the boundary node and the solid node encountered along the $i$-th reticular direction (with coordinates $({\bf x}_s)$):
\begin{equation}
{q_i({\bf x}) = \frac{\sqrt{({\bf x}_b-{\bf x}_L)^2}}{\sqrt{({\bf x}_b-{\bf x}_s)^2}}}\, .
\label{qbouzidi}
\end{equation}      
Finally, for all of the boundary nodes the incoming post-collision distribution functions are bounced-back following the rule:
\begin{equation}
{\begin{cases}
f_{\overline{i}}({\bf x}) = q_i({\bf x})(2q_i({\bf x})+1)f_i({\bf x}) + (1-4q_i^2({\bf x}))f_i({\bf x}-{\bf e}(i)\Delta t) -  q_i({\bf x})(1-2q_i({\bf x}))f_i({\bf x}-2{\bf e}(i)\Delta t) \\ $\hfill if\quad$  q_i({\bf x}) < 0.5 \\

f_{\overline{i}}({\bf x}) = \frac{1}{q_i({\bf x})(2q_i({\bf x})+1)}f_i({\bf x}) + \frac{2q_i({\bf x}) -1}{q_i({\bf x})}f_i({\bf x}) + \frac{1-2q_i({\bf x})}{1+2q_i({\bf x})}f_i({\bf x}-{e}_i\Delta t)  $\hfill if\quad$  q_i({\bf x}) \geq 0.5

\end{cases}}\, ,
\label{bbouzidi}
\end{equation}
where $\overline{i}$ indicates the opposite reticular direction with respect to $i$.

\section{Results and discussion}

\subsection{\textsc{Computational set-up and boundary conditions}}

The deformation of a capsule navigating a microfluidic bifurcation is studied here in terms of the capsule mechanical stiffness Ca, the occlusion ratio $d_p/d$ (ratio between the particle $d_p$ and the main channel $d$ diameters), the bifurcation aperture angle $\theta$ as well as its junction geometry, and the number of flowing capsules. In particular, this systematical analysis is composed of two numerical experiments campaigns: i) the transport of a single capsule and ii) the transport of a train of three capsules into a microfluidic bifurcation. Firstly, the authors assess the transport of an isolated capsule by measuring the relative variation of the capsule surface area $\delta S$ while navigating the microfluidic chip for all of the possible combinations of the geometrical and mechanical parameters involved. The ability of such isolated capsule of selecting a path toward one of the two branches is not considered in this work. As a matter of fact, due to the length of the main channel, it is reasonable to suppose that such capsule would reach the channel midway before entering the junction region. With this rationale, the releasing lateral position is kept constant as 0.5 d. Note that, for small Reynolds number, hydrodynamics do not prescribe any reason to choose one of the two daughter branches. The role of the capsule's inertia is non-negligible only during the interaction with the vasculature. Due to these rationales, the authors do not discuss the path-selection problem since it is random in principle. Indeed, the randomness of nature losts in computations and the selection of one of the two branches may be due to several operating factors like the indexing of Lagrangian mesh triangles in exerted stresses evaluations or Eulerian grid resolution. Then, for a specific occlusion ratio, $d_p/d = 2/3$, the journey of a train of three capsules as a function of Ca and $\theta$ is investigated and characterized in terms of the path traveled and relative stretching. Lastly, this journey is assessed when transporting smaller rigid particles within the train of capsules and analyzing their mutual interaction.  

\begin{figure}[h!]
\centering
\includegraphics[scale=0.25]{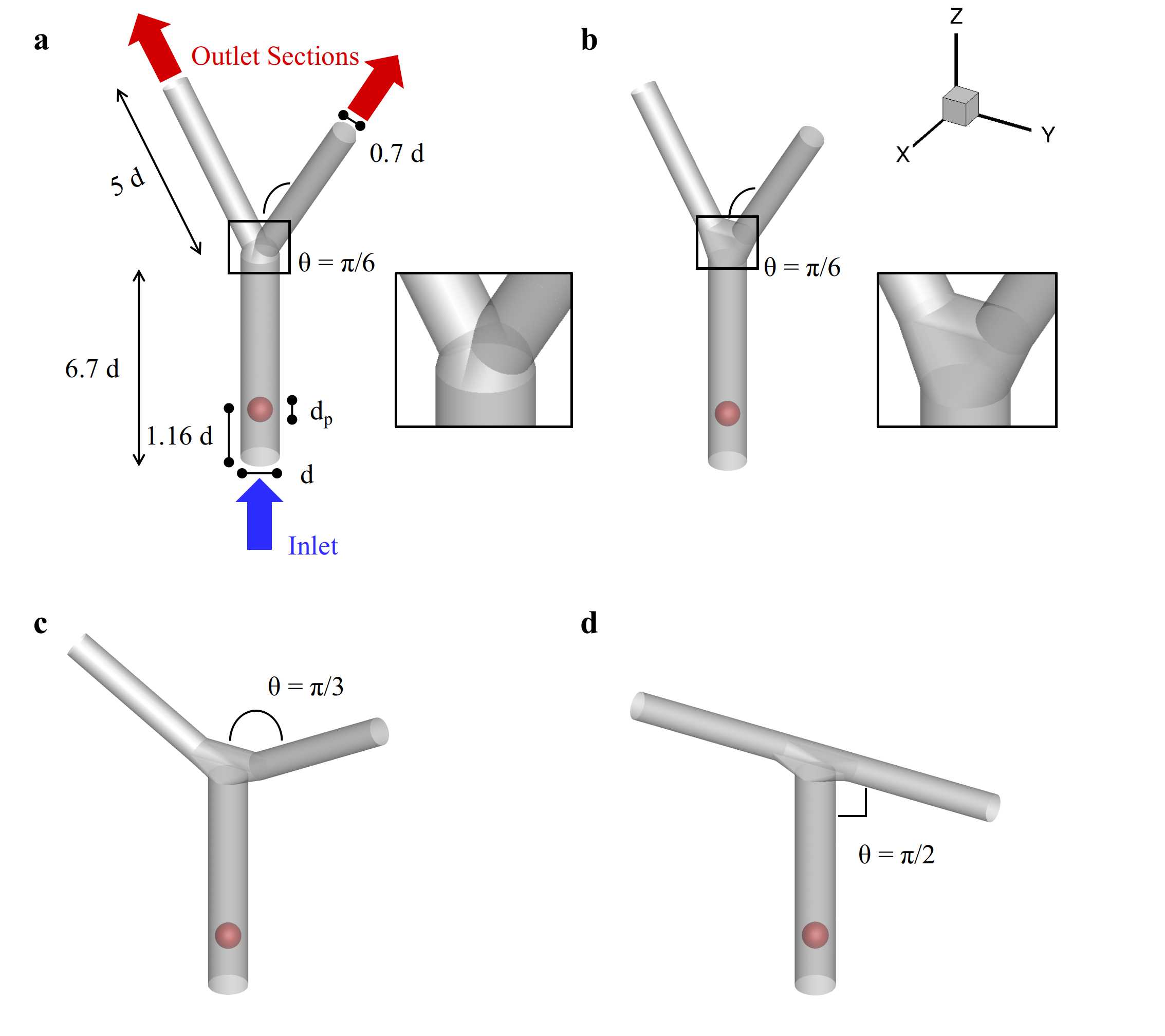}
\caption{{\bf Schematic of the microfluidic bifurcation.} {\bf a.} Sketch of the microfluidic chip along with characteristic dimensions and lengths when considering the sharp junction between the main and the two daughter-branches. ({\bf b, c, d}) Sketch of the microfluidic bifurcations with the smooth junction for $\theta = \pi/6$  ({\bf b}), $\theta = \pi/3$ ({\bf c}), and $\theta = \pi/2$ ({\bf d}).}
\label{schematic}
\end{figure}

A circular cross-section channel with diameter d = 30 $\Delta x$ branches in two daughter ducts with diameter 0.7 d so as to maintain constant the total cross-section (see {\bf Figure.\ref{schematic}a}). The main channel length is 6.7 d while the two branches are 5 d long each. The semi-aperture angle $\theta$ is computed as the angle between the main channel axis and the axis of one of the two branches and corresponds to $\theta =$ $\pi/6$, $\pi/3$, and $\pi/2$. Moreover, two different geometries for the junction of the three channels are considered, a sharp junction and a smoother joint with a trapezoidal vertical section, reported in {\bf Figure.\ref{schematic}}. An initially spherical capsule is placed at 1.16 d from the inlet section (see {\bf Figure.\ref{schematic}a}); 4 different values of the particle diameter are considered, namely, $d_p = \frac{1}{3} d$, $\frac{1}{2} d$, $\frac{2}{3} d$, and $\frac{5}{6} d$. The number of triangular elements $n_L$ discretizing the capsule is 325, 485, 645 and 785 for $d_p/d$ = 1/3, 1/2, 2/3, and 5/6, respectively. As detailed in the computational method section, the mechanical stiffness of such object is regulated by two parameters, the capillary number Ca and the bending resistance Eb. In this work Ca = 0.01, 0.025, 0.05, 0.075 and 0.1 while the bending modulus equals Eb = 0.001. The capsules mass is computed through the ratio between solid and fluid densities, fixed here as $\rho_s/\rho_f = 1.1$.
At the inlet section, the velocity-based Dirichlet conditions are imposed and the Poiseuille parabolic velocity profile with peak velocity $u_{max}$ is adopted, thus, mimicking the velocity profile of an already established flow. At the two outlet sections the Dirichlet pressure-based outlet conditions with $p_{out}=p_{ref}$ are imposed. $u_{max}$ is regulated by the Reynolds number of the flow defined as Re = $\frac{u_{max} d}{\nu}$ = 0.01. The kinematic viscosity $\nu$ depends on the relaxation time as $\nu=c_s^2(\tau -\frac{\Delta t}{2})$ being $\tau = 0.75$ and $\Delta t = 1$~\cite{coclite20204}. Notably, Re = 0.01 correspond to a peak velocity of $u_{max}$ = 0.001 m/s for a 10 $\mu$m particle transported in plain water.
The flow field is initialized with the solution of the stationary flow obtained when no particles or capsule is transported into the bifurcation (see {\bf Supp.Figure.1} and {\bf .2}).

\subsection{\textsc{Transport of an elastic capsule in a microfluidic bifurcation}}

\begin{figure}[h!]
\centering
\includegraphics[scale=0.2]{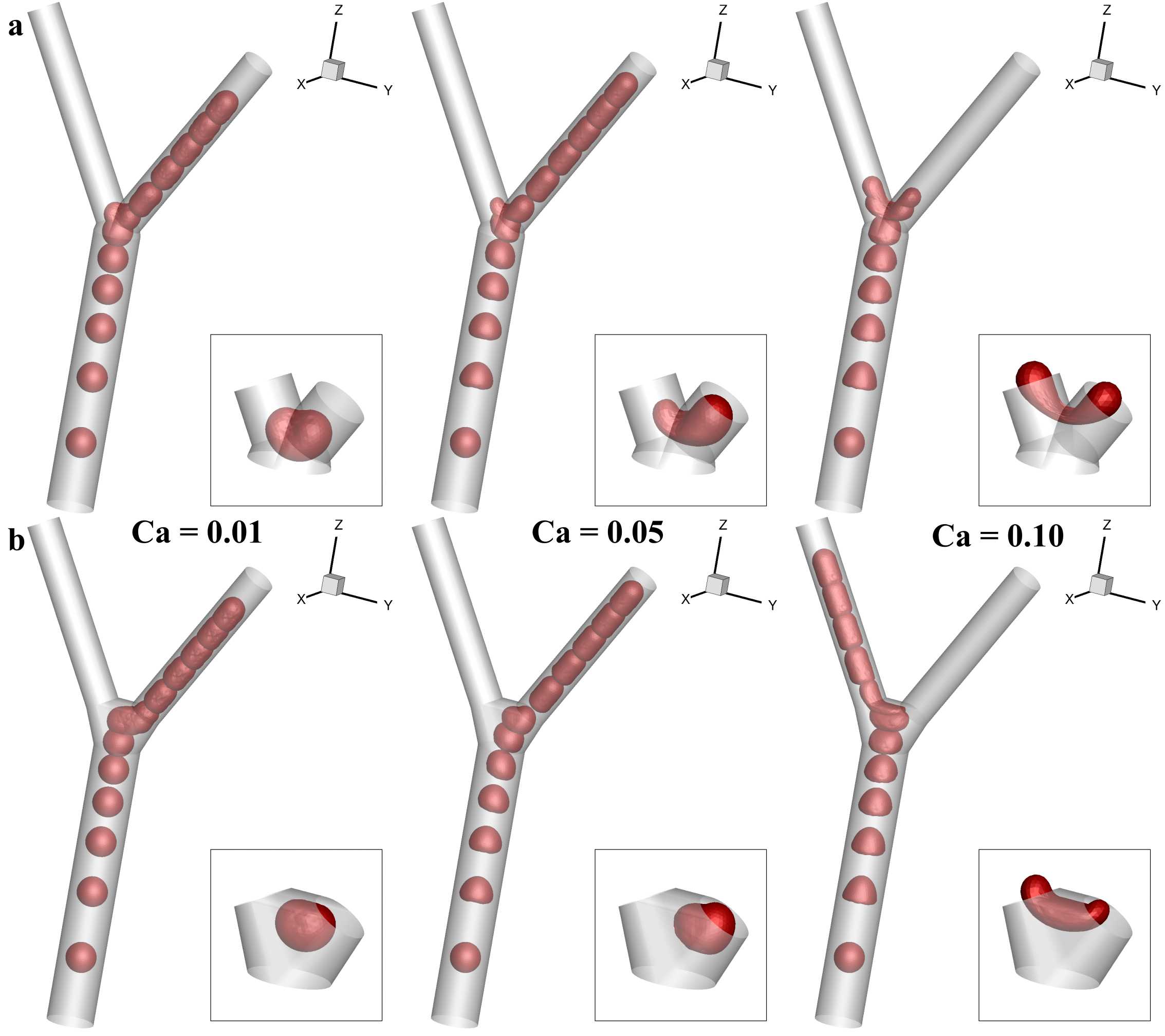}
\caption{{\bf Single capsule crossing the bifurcation for $\theta = \pi/6$ as a function of Ca.} Configurations of a single capsule navigating a bifurcation with the sharp ({\bf a.}) and the smooth junction ({\bf b.}) for Ca = 0.01 (left column), Ca = 0.05 (central column), and Ca = 0.10 (right column). The configurations correspond to $t u_{max}/d$ from 0 to 40 with a step of 3.5. Close-ups refer to impinging instant depending, indeed, by the capsule stiffness.}
\label{Theta30_Conf}
\end{figure}

{\bf The role of the junction geometry.} The transport of the capsule with $d_p/d$ = 2/3 and the analysis of its deformation when crossing the microfluidic chip is firstly considered to compare the abilities of the two junctions proposed in {\bf Figure.\ref{schematic}a} and {\bf .\ref{schematic}b}. Specifically, the ability of the two junctions in facilitating the journey of the microcapsule is assessed considering the probability of large deformations induced by the specific geometry. In this work, a link connecting two vertices of the Lagrangian mesh composing the capsule is considered \textit{damaged} or \textit{broken} when the actual length doubles its value in the stress-free configuration (at this occurrence the computation is stopped). The deformation of a single capsule transported within the bifurcation as a function of the mechanical stiffness for $\theta = \pi/6$ is analyzed. The configurations acquired during the transport of the capsule for Ca = 0.01, 0.05 and 0.1 are documented for the sharp and the smooth junction in {\bf Figure.\ref{Theta30_Conf}a} and in {\bf \ref{Theta30_Conf}b}, respectively. 
The capsule is released for $t u_{max}/d = 0$ and navigates the main channel while deforming under the established flow.
Then, entering the junction region the capsule impinges on the upper wall and, depending on the geometry, it may be stretched and damaged. When capsules hit the cusp in the sharp junction (see close-up in {\bf Figure.\ref{schematic}a}) they squeeze while adhering to the cusp. At this stage, the capsule peripheries are equally distributed in both daughter branches (as documented in the close-ups of {\bf Figure.\ref{Theta30_Conf}a}). Immediately after this phase, the capsule randomly chooses a path to follow selecting one of the two branches and crawling the cusp edge eventually navigating the selected duct. However, for Ca = 0.1 the capsule results \textit{damaged} during the impingement with the cusp edge; in fact, it deforms more and more responding to the symmetry of the flow, which never pushes the particle in one of the two branches.  
\begin{figure}[h!]
\centering
\includegraphics[scale=0.225]{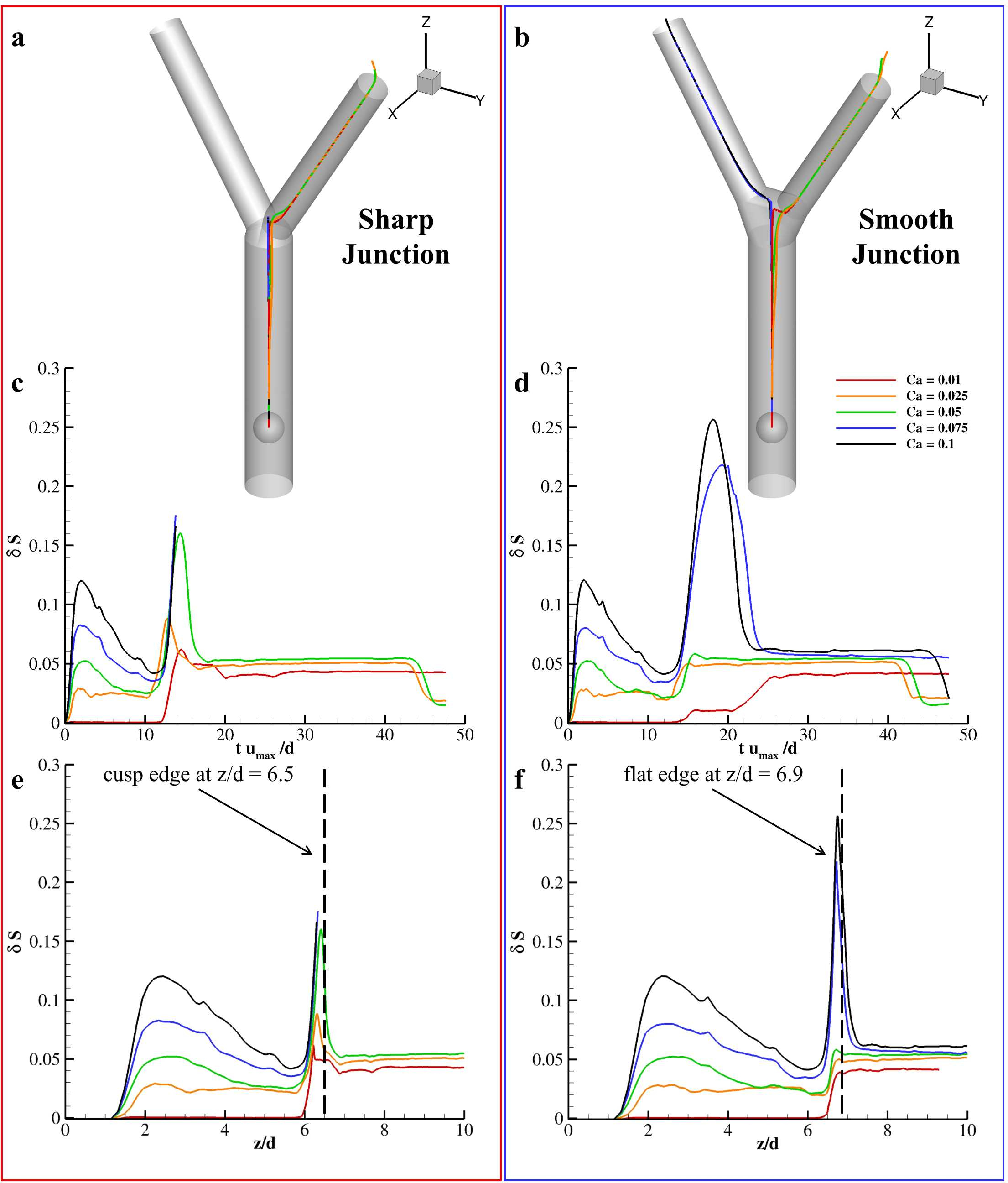}
\caption{{\bf Trajectory and relative stretching of the single capsule crossing the bifurcation with $\theta = \pi/6$ as a function of the mechanical stiffness.} ({\bf a}, {\bf b}) Path traveled by a capsule into the microfluidic bifurcation for $\theta = \pi/6$ and Ca = 0.01, 0.025, 0.05, 0.075, and 0.1 with the sharp ({\bf a}) and the new ({\bf b}) junction. Distribution of $\delta S$ over time obtained with the sharp ({\bf c}) and the new ({\bf d}) junction. Distribution of $\delta S$ over the vertical coordinate $z/d$ for the sharp ({\bf e}) and the new ({\bf f}) junction. The dashed lines correspond to the vertical coordinate of the junction edge.}  
\label{Theta30_Traj}
\end{figure}
This picture results quite different when considering the smoother joint presented in {\bf Figure.\ref{schematic}b}. Indeed, by adopting a junction with a trapezoidal section, the capsule will impinge and adhere to a flat wall. Then, again, it deforms responding to the symmetry of the flow, and crawls over this flat wall until it is dragged into one of the two smaller vessels (see {\bf Figure.\ref{Theta30_Conf}b}). Interestingly, with this smooth junction, the capsule can overcome the bifurcation and travel one of the two branches for all of the investigated values of Ca. However, at the impinging instant, all of the capsules transported within the smooth junction present smaller deformation, as reported in {\bf Figure.\ref{Theta30_Conf}b}. The trajectory of the transported capsules as a function of the stiffness is plotted {\bf Figure.\ref{Theta30_Traj}} along with the distribution of the relative stretching $\delta S$. The relative stretching is defined as the ratio $\frac{S(t)-S_0}{S_0}$ where $S(t)$ and $S_0$ are the current and the initial capsule surface area. As discussed, the capsules stiffness does not seem to play any role in the choice of which branch will be traveled after overcoming the junction. Moreover, the distribution in time of the relative stretching demonstrates how for a fixed capillary number the geometry of the junction may induce larger or smaller deformation along with the occurrence of the capsule membrane degradation. For Ca = 0.05, for example, the peak value of $\delta S$ is 0.161 (for $t u_{max/d}$ = 14.28) for the sharp junction and 0.058 (for $t u_{max/d}$ = 15.70) for the smooth junction; for Ca = 0.075, the capsule results damaged when using the sharp junction while with the smooth junction the peak value of $\delta S$ registered corresponds to 0.22 at $t u_{max}/d$ = 19.04 (see {\bf Figure.\ref{Theta30_Traj}c} and {\bf .\ref{Theta30_Traj}d}). The time instant in which the peak value of $\delta S$ is registered depends, indeed, on the capsule's mechanical stiffness. In fact, on one side, softer capsules spend a part of the momentum transferred by the fluid for deforming their surface; on the other side, stiffer capsules are gradually deformed by the flow and spend more time acquiring the needed deformation to navigate a daughter vessel. On the contrary, the position along the vertical coordinate in which the most of the deformation is performed, so that, in which the peak of $\delta S$ is registered, depends on the peculiar flow patterns obtained and not only on Ca. However, when considering the sharp junction, the position in which those peaks are registered varies with Ca, as documented in {\bf Figure.\ref{Theta30_Traj}e}. While, this phenomenon becomes negligible whit the smooth junction (see {\bf Figure.\ref{Theta30_Traj}f}). The peak values registered when using the sharp junction correspond to 6.22, 6.30, and 6.39 $z/d$ for Ca = 0.01, 0.025, and 0.05, respectively. While such peaks are measured at 6.70, 6.72, 6.72, 6.72, and 6.74 $z/d$ for Ca = 0.01, 0.025, 0.05, 0.075, and 0.1 whit the smooth junction. 
\begin{figure}[h!]
\centering
\includegraphics[scale=0.15]{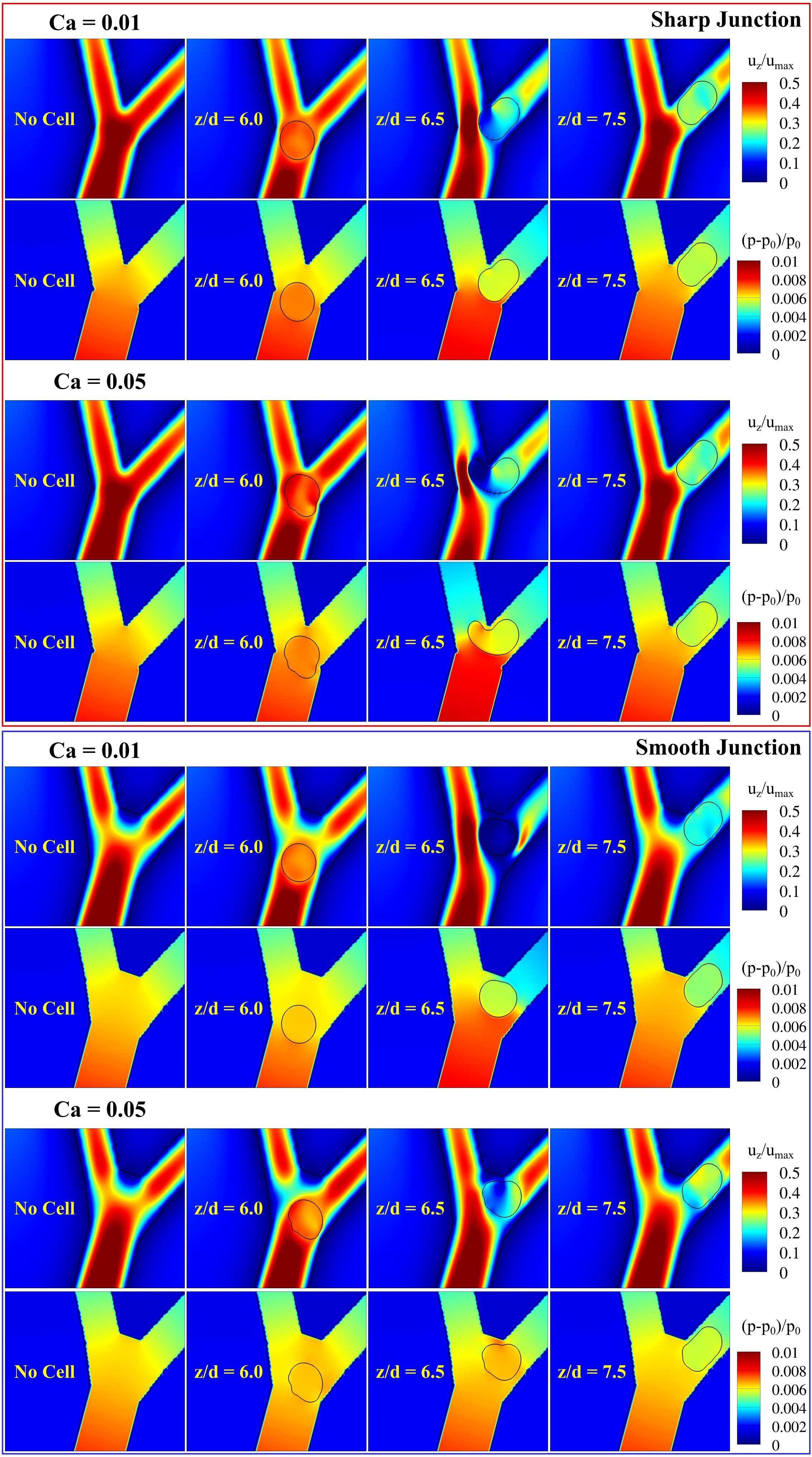}
\caption{{\bf Flow patterns induced by the presence of the transported capsule during the impinging with the junction edge for $\theta = \pi/6$ and Ca = 0.01, 0.05 obtained with both junctions.}}
\label{Theta30_FP}
\end{figure}
Such difference is ascribable to the peculiar flow patterns obtained with the two. The smooth junction presents a flat edge close to which the fluid is resting like in a sort of stagnation region (see no-cell contour plots in {\bf Figure.\ref{Theta30_FP}}). This stagnation region slows down the capsules letting them to acquire the needed deformation to navigate the secondary branches. Indeed, also the fluid flow is perturbed by the capsule presence in two different fashions when considering the two junctions. Such perturbations are measured here by seeing at the vertical component of the velocity field $u_z/u_{max}$ as well as the relative pressure $(p-p_0)/p_0$, being $p_0$ the outflow pressure (see {\bf Figure.\ref{Theta30_FP}}). When the capsule impinges on the cusp edge of the sharp junction it squeezes adhering to it and implying a partial occlusion of both branches whose extension depends on Ca. Softer capsules, indeed, occupy larger regions of both branches during this phase (see the plots for $z/d = 6.5$ obtained with the sharp junction in {\bf Figure.\ref{Theta30_FP}}). Interestingly, such partial occlusion is negligible when using the smooth junction and the capsules are able to navigate the vasculature without spreading into both branches and perturbing the flow in both secondary channels (see the lower panel in {\bf Figure.\ref{Theta30_FP}}).

\begin{figure}[h!]
\centering
\includegraphics[scale=0.125]{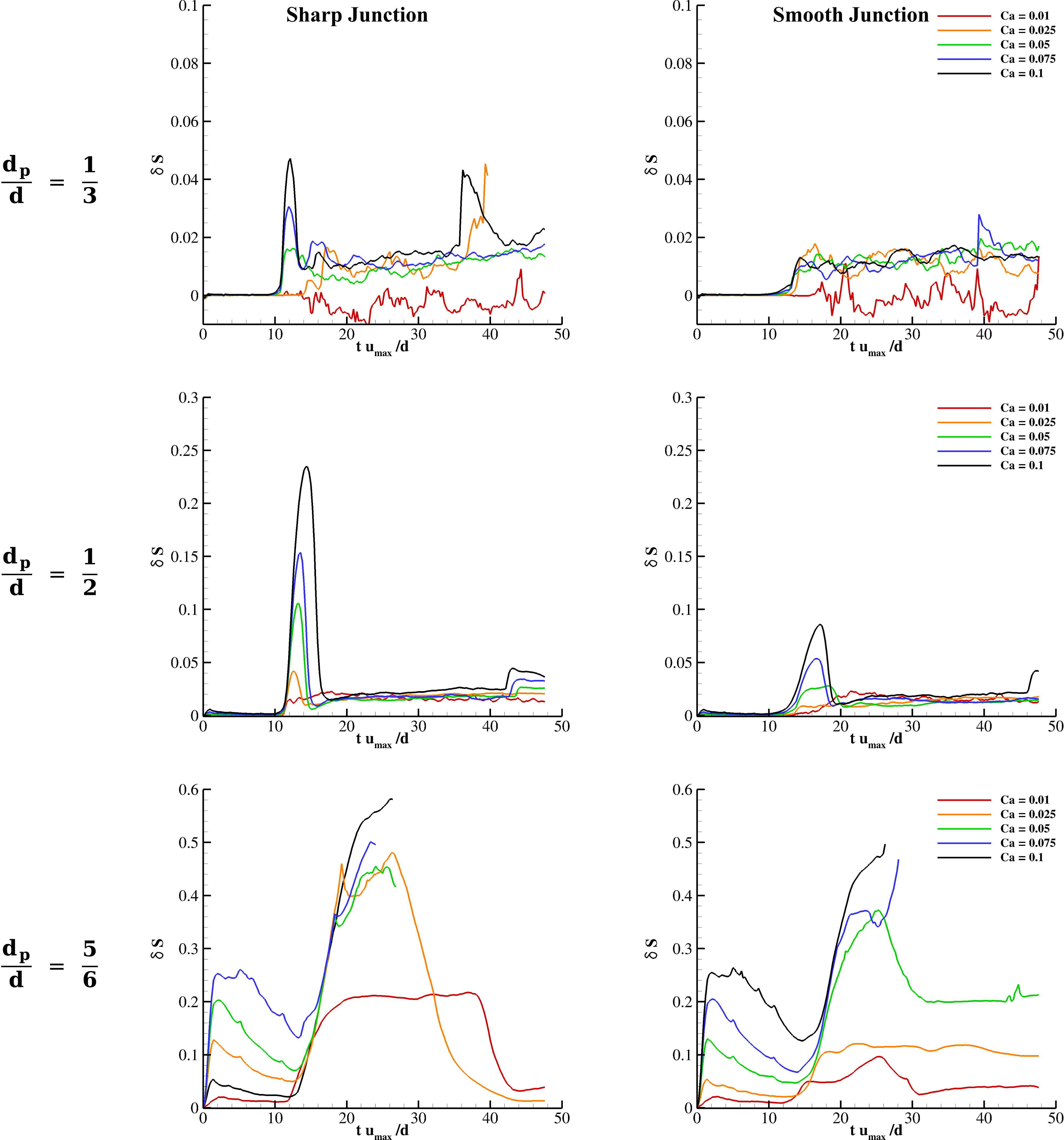}
\caption{{\bf Stretching of a single capsule transported in the bifurcation for $\theta = \pi/6$ as a function of $d_p/d$ and Ca for both junctions.}}
\label{Theta30_BR}
\end{figure}
The ability of the two junctions in facilitating the journey of a microcapsule is further analyzed in terms of the occlusion ratio $d_p/d$. Specifically, the particle diameter is varied from 1/3 d to 5/6 d with a 1/6 d step. The distributions of $\delta S$ over time are documented in {\bf Figure.\ref{Theta30_BR}} for $d_p/d$ = 1/3, 1/2, and 5/6. As expected, for $d_p/d$ = 1/3 the capsule experience almost negligible deformations (smaller than 0.05) for all of the investigated capillary numbers. However, the sharp junction still implies much stronger deformations with respect to the new one. 
For $d_p/d$ = 1/2 deformations are valuable and the peak values registered are almost doubled when considering the sharp over the smooth junction. Specifically, the peak value of $\delta S$ corresponds to 0.106, 0.154, and 0.235 (taken at $t u_{max/d}$ = 13.33, 13.57, and 14.28) for Ca = 0.05, 0.075 and 0.1, respectively, when using the sharp joint; while, it equals 0.025, 0.054, and 0.086 (taken at $t u_{max/d}$ = 16.03, 16.67, and 17.14) for Ca = 0.05, 0.075 and 0.1, respectively, when using the smooth junction (see plots in the middle row of {\bf Figure.\ref{Theta30_BR}}). Notably, for $d_p/d$ = 1/2 no capsule results {\it damaged} during the journey. On the contrary, for the largest occlusion ratio the capsule results {\it damaged} for Ca = 0.05, 0.075 and 0.1 for the sharp junction and for Ca = 0.075 and 0.1 for the new joint. Indeed, the measured peak values of $\delta S$ increase with $d_p/d$ due to needed squeezing for passing the bifurcation. From this comparison, we conclude that smoother geometries for the junctions should be preferred in the design phase of a microfluidic chip against a net joining of the two daughter branches onto the main channel and from now on we will further analyze the deformation of capsule only considering the smooth junction.

\begin{figure}[h!]
\centering
\includegraphics[scale=0.2]{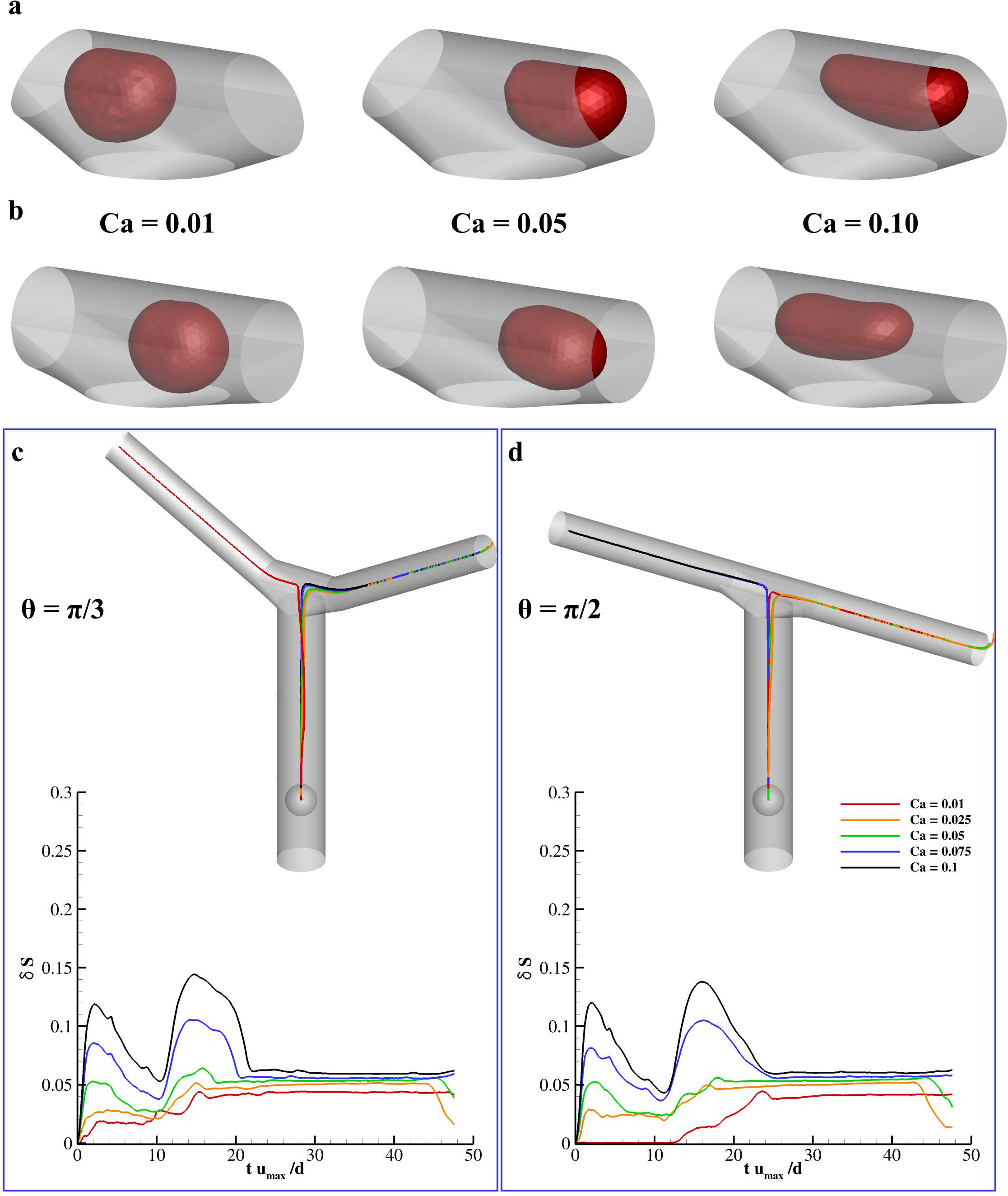}
\caption{{\bf Single capsule transported in the bifurcation for $\theta = \pi/3$ and $\pi/2$} ({\bf a, b}) Configuration of the capsule at the impinging instant for Ca = 0.01 (left column), 0.05 (central column), and 0.1 (right column) for $\theta = \pi/3$ ({\bf a}) and $\pi/2$ ({\bf b}). ({\bf c, d}) Trajectory and relative stretching of the capsule navigating the bifurcation for $\theta = \pi/3$ ({\bf c}) and $\pi/2$ ({\bf d}) as a function of Ca.}
\label{Theta60-90}
\end{figure}

{\bf The role of the bifurcation aperture angle.} The deformation of a capsule navigating a microfluidic bifurcation is further assessed when varying the semi-aperture angle of the two branches with respect to the main channel (see {\bf Figure.\ref{schematic}c} and {\bf .\ref{schematic}d}). For $\theta = \pi/3$ and $\pi/2$ the deformation experienced by the capsule at the impinging instant is smaller than that corresponding to $\theta = \pi/6$ (see {\bf Figure.\ref{Theta60-90}a} and {\bf .\ref{Theta60-90}b}).
Indeed, for $\theta = \pi/6$ the flow velocity has a large component along z in the junction region dramatically pushing the capsule against the upper wall of the junction (see  {\bf Supp.Figure.1}); however, for larger $\theta$ the geometry of the bifurcation implies an abrupt change in the flow direction that results in a small ($\theta = \pi/3$) or null ($\theta = \pi/2$) component along z into the junction region and the capsule is essentially dragged by inertia against the junction boundary (see  {\bf Supp.Figure.1}). 
The deformed configuration of the capsule as a function of Ca at the impinging instant is documented in {\bf Figure.\ref{Theta60-90}a} and {\bf \ref{Theta60-90}b} for $\theta = \pi/3$ and $\pi/2$, respectively. Note that, for Ca = 0.01, the capsule does not touch the junction wall for both values of $\theta$. Moreover, also in these two cases the path of such capsules do not seem to be influenced by Ca, as demonstrated in {\bf Figure.\ref{Theta60-90}c} and {\bf .\ref{Theta60-90}d}. The distribution of the relative surface stretching as a function of the stiffness is reported in {\bf Figure.\ref{Theta60-90}}. Quantitatively, the peak values of $\delta S$ registered for Ca = 0.075 are 0.105 (at $t u_{max}/d$ = 14.3) and 0.101 (at $t u_{max}/d$ = 16.2) for $\theta = \pi/3$ and $\pi/2$, respectively, corresponding in both cases at about an half of the peak values obtained for $\theta = \pi/6$. 

\begin{table}
\centering
\begin{tabular}{||c | l | c | c | c | c||} 
\hline
 & & $\theta = \pi/6$ Sharp Jun. & $\theta = \pi/6$ New Jun. & $\theta = \pi/3$ & $\theta = \pi/2$ \\ [0.5ex] 
\hline\hline
 & Ca = 0.01  & $\leq 0.002$ & $\leq 0.002$ & $\leq 0.002$ & $\leq 0.002$ \\ 
 & Ca = 0.025 & $\leq 0.002$ & $\leq 0.002$ & $\leq 0.002$ & $\leq 0.002$ \\
$d_p/d$ = 1/3 & Ca = 0.05  & 0.016 & $\leq 0.002$ & $\leq 0.002$ & $\leq 0.002$ \\
 & Ca = 0.075 & 0.035 & $\leq 0.002$ & $\leq 0.002$ & $\leq 0.002$ \\
 & Ca = 0.1   & 0.047 & $\leq 0.002$ & $\leq 0.002$ & $\leq 0.002$ \\
\hline
              & Ca = 0.01  & 0.017 & 0.006 & $\leq 0.002$ & $\leq 0.002$ \\ 
              & Ca = 0.025 & 0.042 & 0.009 & 0.007 & 0.005 \\
$d_p/d$ = 1/2 & Ca = 0.05  & 0.106 & 0.025 & 0.021 & 0.011 \\
              & Ca = 0.075 & 0.154 & 0.054 & 0.023 & 0.017 \\
              & Ca = 0.1   & 0.235 & 0.086 & 0.028 & 0.024 \\
\hline
              & Ca = 0.01  & 0.062   & 0.010 & 0.043 & 0.034 \\ 
              & Ca = 0.025 & 0.088   & 0.049 & 0.051 & 0.049 \\
$d_p/d$ = 2/3 & Ca = 0.05  & 0.161   & 0.058 & 0.062 & 0.056 \\
              & Ca = 0.075 & Damaged & 0.220 & 0.105 & 0.101 \\
              & Ca = 0.1 	& Damaged & 0.257 & 0.143 & 0.137 \\
\hline
               & Ca = 0.01  & 0.212   & 0.096 	  & 0.092 & 0.087 \\ 
               & Ca = 0.025 & 0.459   & 0.120 	  & 0.126 & 0.111 \\
 $d_p/d$ = 5/6 & Ca = 0.05  & Damaged & 0.372	  & 0.159 & 0.158 \\
               & Ca = 0.075 & Damaged & Damaged & 0.228 & 0.224 \\
               & Ca = 0.1   & Damaged & Damaged & 0.317 & 0.302 \\
\hline
\end{tabular}
\caption{Measured surface strain $\delta S$ achieved by the capsule at the impinging instant as a function of $d_p/d$ and Ca for $\theta$ = $\pi/6$ (both junction geometries), $\pi/3$ and $\pi/2$.}
\label{Tab.dS}
\end{table}

The deformation of the capsule navigating the bifurcation for different occlusion rate is provided also for $\theta$ = $\pi/3$ and $\pi/2$. The distributions of $\delta S$ over time as a function of $d_p/d$ and Ca are collected in {\bf Supp.Figure.3}. Interestingly, for larger $\theta$ almost negligible deformations (smaller than 0.05) are registered for $d_p/d \leq  1/2$. For larger occlusion ratio, the trends enlighten for $\theta = \pi/6$ are confirmed for large semi-aperture angles. For completeness, the measured values of $\delta S$ at the impinging instant are documented in {\bf Table.\ref{Tab.dS}} for all of the combinations of the investigated parameters.  

\subsection{\textsc{Transport of multiple capsules in a bifurcation}}

\begin{figure}[h!]
\centering
\includegraphics[scale=0.2]{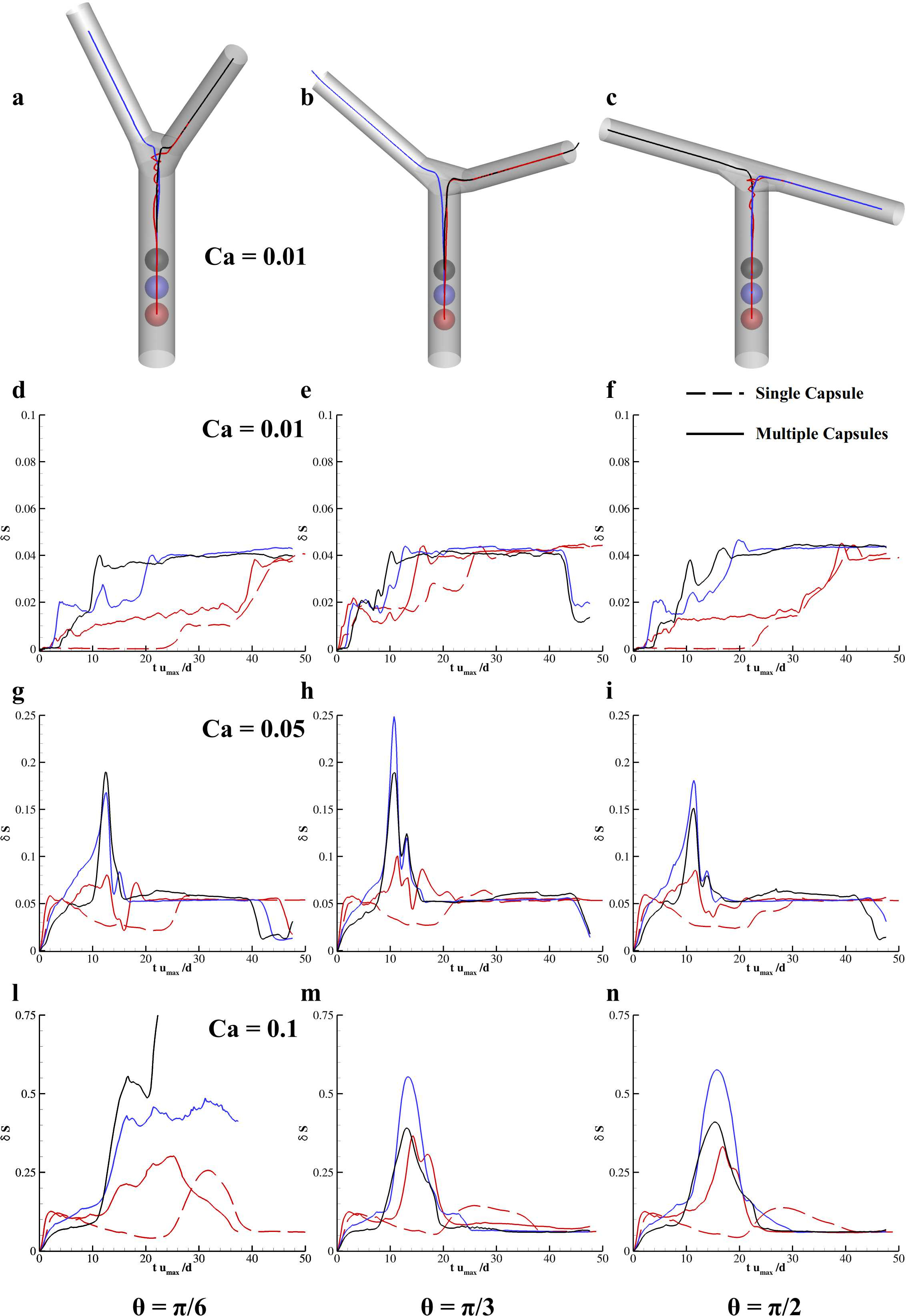}
\caption{{\bf Train of three capsules navigating a microfluidic bifurcation.} ({\bf a, b, c}) Trajectory of the capsules transported into the bifurcation for $\theta = \pi/6$ ({\bf a}), $\pi/3$ ({\bf b}), and $\pi/2$ ({\bf c}) and Ca = 0.01. ({\bf d, e, f}) Distribution of the relative stretching of the capsules for $\theta = \pi/6$ ({\bf d}), $\pi/3$ ({\bf e}), and $\pi/2$ ({\bf f}) and Ca = 0.01. ({\bf h, g, i}) Distribution of the relative stretching of the capsules for $\theta = \pi/6$ ({\bf h}), $\pi/3$ ({\bf g}), and $\pi/2$ ({\bf i}) and Ca = 0.05. ({\bf l, m, n}) Distribution of the relative stretching of the capsules for $\theta = \pi/6$ ({\bf l}), $\pi/3$ ({\bf m}), and $\pi/2$ ({\bf n}) and Ca = 0.1. Colors refer to the specific capsules.}
\label{Train_AllSoft}
\end{figure}

A train of three initially spherical capsules with occlusion ratio $d_p/d$ = 2/3 is considered to further analyze the deformation of capsules crossing the bifurcation. Specifically, two additional capsules are placed on top of the first one, kept in the same position as before, namely 1.16 d from the inlet section, with a mutual distance between the three equal to 0.85 d (see {\bf Figure.\ref{Train_AllSoft}}). 
Interestingly, when transporting multiple capsules into the bifurcation the trajectory patterns complicate. When crossing the bifurcation all particles may navigate one of the two branches in the same fashion as before. However, as already observed, the total cross-section of the main channel is conserved across the bifurcation so that the capsules have to squeeze to enter into a daughter branch after crossing the junction region. Therefore, the established flow results to be partially occluded by the presence of a capsule within one of the two smaller channels, and the last particle (colored in red in {\bf Figure.\ref{Train_AllSoft}}) cannot be easily dragged across the junction zone. For Ca = 0.01 this phenomenon is documented in terms of traveled path in {\bf Figure.\ref{Train_AllSoft}a--.\ref{Train_AllSoft}c} for $\theta = \pi/6$, $\pi/3$, and $\pi/2$ while the entire capsules' journey is reported in {\bf Supp.movie.1}--{\bf .3}. Moreover, the distribution of the relative stretching registered for the three capsules is compared with the one obtained for the transport of a single capsule used as reference. Large stretching is observed almost for all of the investigated combinations of $\theta$ and Ca (see {\bf Figure.\ref{Train_AllSoft}d--.\ref{Train_AllSoft}n}) while for Ca = 0.1 and $\theta = \pi/6$ the degradation of both upper capsules is registered. Indeed, the relative stretching distributions indicate that for $\theta = \pi/3$ and $\pi/2$ the second capsule squeezes more than the other two for all Ca due to the presence of the first capsule and the relative obstruction seen when crossing the junction zone (see {\bf Supp.Movie.4-.6}).

\begin{figure}[h!]
\centering
\includegraphics[scale=0.2]{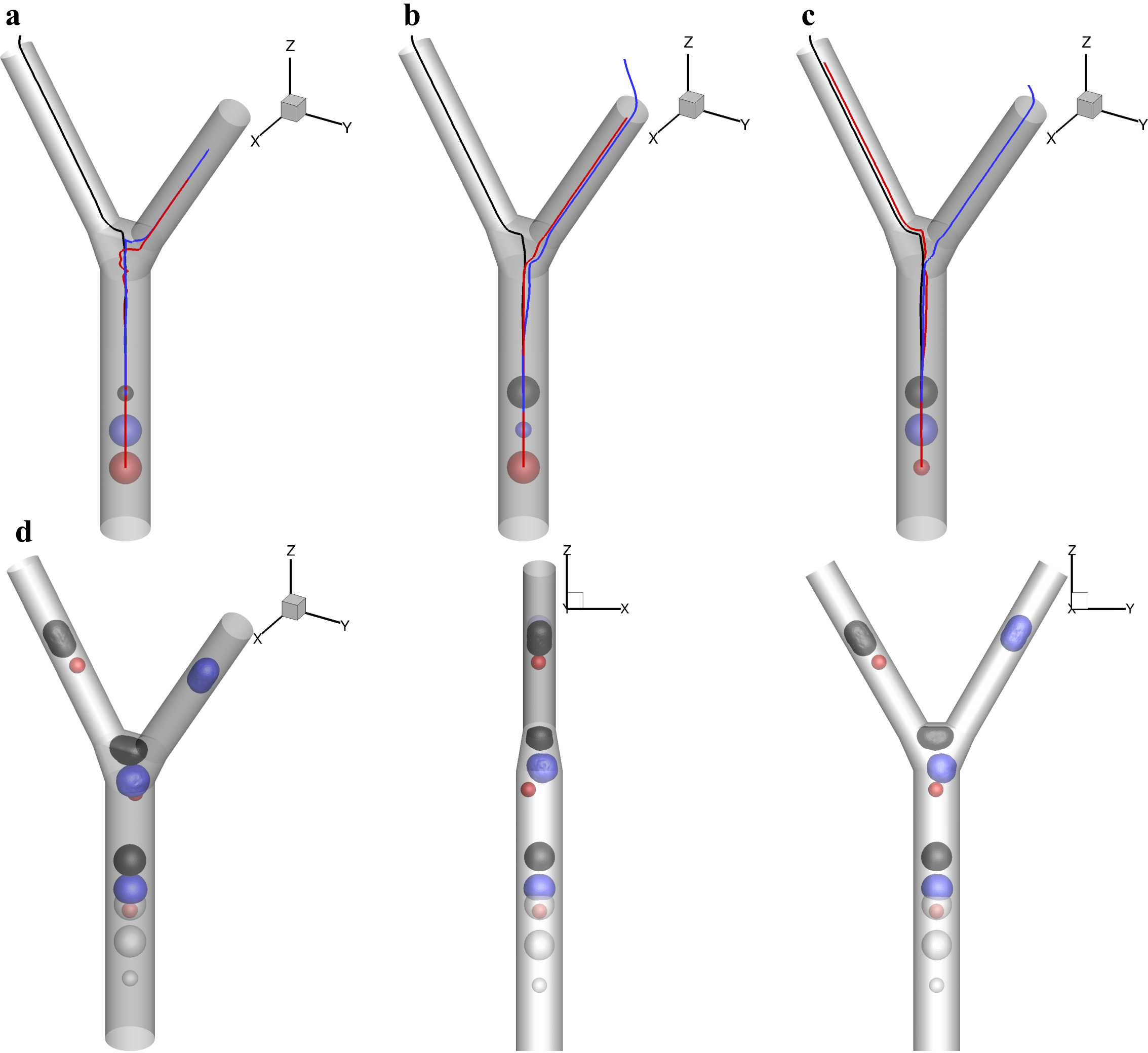}
\caption{{\bf Train of rigid and soft capsules navigating the bifurcation for $\theta = \pi/6$.} ({\bf a, b, c}) Trajectory of the capsules transported into the bifurcation for $\theta = \pi/6$ when considering a rigid capsule in the top ({\bf a}), middle ({\bf b}), and bottom ({\bf c}) releasing position. {\bf d.} Configuration of the transported capsules for $\theta = \pi/6$ and with the rigid capsule in the bottom releasing position from three different point of views for $t u_{max}/d$ = 0, 5, 15, 40.}
\label{Train_1Rig_ALL}
\end{figure}
The transport of microvesicles is assessed also when considering rigid objects and their interaction with deformable membranes for $\theta = \pi/6$ and Ca = 0.01. In particular, three more configurations have been proposed by placing a rigid plain sphere with the same density of soft capsules and diameter $\frac{1}{2} d_p$ in place of a deformable capsule (a releasing position per time). All of the initial conditions are reported in {\bf Figure.\ref{Train_1Rig_ALL}a}.
\begin{figure}[h!]
\centering
\includegraphics[scale=0.2]{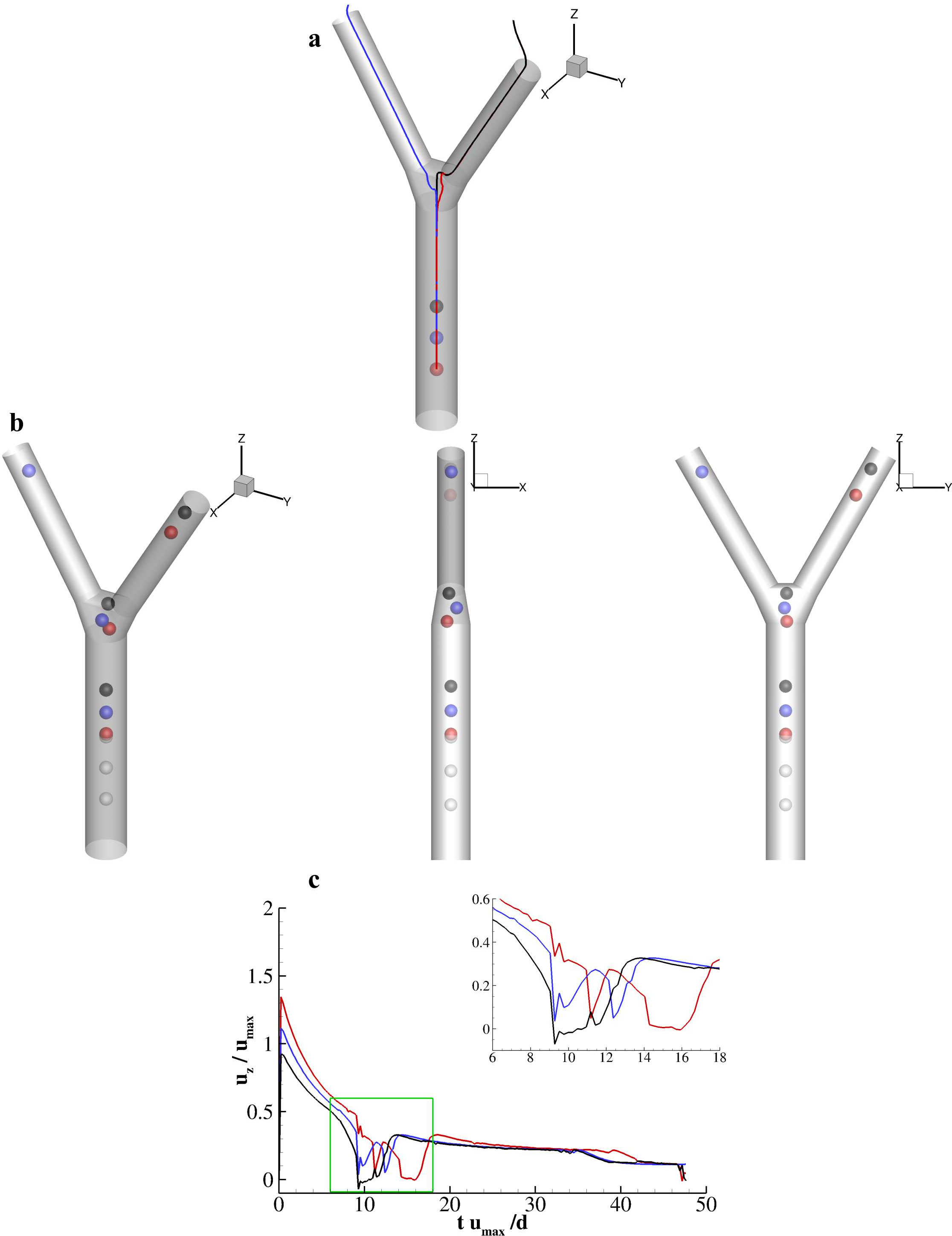}
\caption{{\bf Train of three rigid capsules navigating the bifurcation for $\theta = \pi/6$.} {\bf a.} Trajectory of the capsules transported into the bifurcation. {\bf b.} Capsules configurations during the transport from three different point of views for $t u_{max}/d$ = 0, 5, 10, 40. {\bf c.} Distribution of the z component of the center of mass velocity for the three particle with a close-up corresponding to the crossing of the junction region.}
\label{Train_3Rig}
\end{figure}
The transport of such objects along with their deformation and trajectories is complicated again due to two phenomena: i) rigid objects would travel the bifurcation faster than soft capsules since they are not subject to any deformation crossing the junction region; ii) due to their size, rigid particles would tend to avoid direct interaction with deformable membranes by moving laterally and continuing to follow flow streamlines. The registered trajectories of these three soft or rigid capsules are reported in {\bf Figure.\ref{Train_1Rig_ALL}a--.\ref{Train_1Rig_ALL}c} (see also {\bf Supp.Movie.7--.9}). When the rigid particle is in the upper releasing position it will travel across the bifurcation almost unperturbed by the two following larger capsules (see {\bf Figure.\ref{Train_1Rig_ALL}a}). Note that, again the last capsule (colored in red) follows the path chosen by the soft particle above (colored in blue) due to the occurrence of the flow obstruction and consequent capillary suction operated by the latter (see {\bf Supp.Movie.7}). On the contrary, when placing the rigid sphere in the bottom releasing position (see {\bf Figure.\ref{Train_1Rig_ALL}c}), the latter need to flow laterally to avoid contact with the two larger membranes in the main channel and then moves in a dislodged line with respect to the duct axis trying into the selected daughter-branch. The capsules configurations during these two moments are depicted in {\bf Figure.\ref{Train_1Rig_ALL}d}, the small rigid sphere tends to marginate due to the interaction with larger objects. This phenomenon is enhanced when transporting three rigid particles within the microchip (see {\bf Figure.\ref{Train_3Rig}}). Such rigid particles do not interact between them during the journey dislodging one with respect to the others to find the smoother and more undisturbed path to overcome the bifurcation. For completeness, also in this last case, the three particles' journey is reported in {\bf Supp.Movie.10}. Moreover, the distribution of the streamwise component of the center of mass velocity for the three particles is documented in {\bf Figure.\ref{Train_3Rig}c}. The time interval in which the particle traverse the junction region is enlightened into a close-up. The upper particle impinges against the junction edge at 9.28 $t u_{max} /d$ (black line) and crawls on the right branch. In the meantime, the immediately following particle (colored in blue) selects the left branch thus avoiding any interaction with particles or with the vasculature. For the same reason, the last particle (colored in red) decides to follow the black particle that is already sufficiently far.

\section*{Conclusions and future work}

The deformation of inertial capsules navigating a microfluidic bifurcation as a function of their stiffness and size is critically analyzed in this paper. The incompressible Navier-Stokes equation is modeled through a lattice Boltzmann scheme in which body forces account for immersed geometries. The latter is treated with two different approaches; on one side, no-slip boundary conditions are imposed on moving deformable or rigid geometries through the dynamic-Immersed Boundary method, while, on the other side, on fixed immersed geometries the Bouzidi-Firdaouss-Lallemand second-order bounce back technique is adopted. As per the external boundaries, Dirichlet boundary conditions are imposed with the Zou-He formulation.

The proposed model was adopted to systematically analyze the transport of micrometric capsules when transported in a microfluidic bifurcation as a function of their mechanical stiffness, size, bifurcation aperture angle, main-to-daughter channels junction geometry, and the number of traveling particles. In particular, initially spherical capsules with different mechanical stiffness (Ca = 0.01, 0.025, 0.05, 0.075, and 0.1) and different size ($d_p/d =$ 1/3, 1/2, 2/3, and 5/6) are transported one per time within a bifurcation with semi-aperture $\theta = \pi/6$. This bifurcation consists of the main channel that splits into two daughter ducts. Such splitting is performed with two junctions: a sharp joining of the three channels and a more smooth joint with a trapezoidal vertical section. For $d_p/d <$ 2/3 the smooth junction is found to not injure the membranes transported within thanks to the flat edge on top of the junction connecting the two branches. On the contrary, the sharp junction is found to imply strong deformations on the navigating capsules. Large particle ($d_p/d$ = 5/6) are damaged with both junctions for Ca $>$ 0.05. Moreover, it is demonstrated that for larger aperture angles ($\theta =$ $\pi/3$ and $\pi/2$) smaller deformations are registered for all particles size. The measured distribution of the relative stretching during the transport is documented for all apertures as a function of Ca and $d_p/d$ demonstrating that small particles ($d_p/d$ = 1/3) exert almost null deformations while larger capsules ($d_p/d \geq $ 2/3) may be damaged also with the smoother junction for large values of Ca. 
Then, the transport of a train of three capsules into the bifurcation is assessed for $\theta = \pi/6$ as a function of Ca. Interestingly, when multiple capsules navigate the bifurcation the particle released into the lowest releasing position deal with the flow obstruction due to the presence of the other two membranes. After one of the two obstructing capsules reach the bifurcation peripheries, the flow is restored and the last particle is eventually dragged into a daughter branch. Lastly, those three deformable capsules are substituted with a rigid smaller particle one per time. For the specific set of investigated parameters, smaller rigid particles are found able to dislodge laterally in order to avoid direct interactions with larger capsules.

Collectively, these data provide specific insights for the scientific community investing efforts in the rational design of microfluidic chips for the assessment of particle dynamics and cell sorting in human microcirculation. The spanned region of the phase-space determined by the number of the investigated parameters deepens the knowledge of the fundamental physics behind the prototypical problem of a particle navigating a bifurcation.

\section*{Acknowledgments}
AC acknowledges the project ``Research for Innovation'' (REFIN) - POR Puglia FESR FSE 2014-2020 - Asse X - Azione 10.4 (Grant No. CUP - D94120001410008). AC and TP are members of Gruppo Nazionale per il Calcolo Scientifico (GNCS) of Istituto Nazionale di Alta Matematica (INdAM). This research has been partially supported by the  Research Project of National Relevance ``Multiscale Innovative Materials and Structures'' granted by the Italian Ministry of Education, University and Research (MIUR Prin 2017, project code 2017J4EAYB and the Italian Ministry of Education, University and Research under the Programme Department of Excellence Legge 232/2016 (Grant No. CUP - D94I18000260001). 

\section*{Competing Interests}
The authors declare no competing interests.

\end{document}